\documentclass[useAMS,usenatbib]{mn2e}
\usepackage{graphicx}
\usepackage{xcolor}
\usepackage{amsmath}
\usepackage{url}
\usepackage{color}

\title[]{ `Zwicky's Nonet': a compact  merging ensemble  of  nine galaxies and   4C~35.06, a peculiar radio
galaxy  with dancing  radio jets}
\author[Biju et al.]{K.G. Biju$^{1,5}$\thanks{kgbiju42@gmail.com}, 
Joydeep Bagchi$^{2}$\thanks{joydeep@iucaa.in}, Ishwara-Chandra C.H.$^{3}$,
M. Pandey-Pommier$^{4}$,  \newauthor
Joe Jacob$^{5}$, M.K. Patil$^{6}$, Sunil Kumar P.$^{7}$, Mahadev Pandge$^{8}$\thanks{DST INSPIRE Faculty}, 
\newauthor Pratik Dabhade$^{2}$\thanks{Research Fellow, Indo-French Centre for the Promotion of 
Advanced Research (IFCPAR)/(CEFIPRA)}, Madhuri Gaikwad$^{3}$, Samir Dhurde$^{2}$,  
Sheelu Abraham$^{2}$, M. Vivek$^{9}$, \newauthor Ashish A. Mahabal$^{10}$, 
S. G. Djorgovski$^{10}$
\\
$^1$ WMO Arts \& Science college, Muttil, Wayanad, Kerala, 673122, India\\
$^{2}$ Inter University Centre for Astronomy and  Astrophysics,(IUCAA), Pune University Campus, Post
Bag 4, Pune 411007, India \\
$^3$ National Centre for Radio Astrophysics, TIFR, Post Bag  No. 3, Ganeshkhind, Pune, 411007, India\\
$^{4}$Centre de Recherche Astrophysique de Lyon- Observatoire de Lyon, 9 avenue Charles Andr\'e, \\Saint-Genis Laval, F-69230, France\\
$^5$ Newman College, Thodupuzha, Kerala, 685584, India\\
$^6$ School of Physical Sciences, Swami Ramanand Teerth Marathwada University, Nanded, 431606, India\\
$^7$ P.T.M.Govt. College, Perinthalmanna, Kerala, India\\
$^8$ Department of Physics, Dayanand Science College, Latur, 413512, India \\
$^9$ Department of Physics and Astronomy, University of Utah,Salt Lake City, UT 84112, USA \\
$^{10}$ California Institute of Technology, 1200 E California B1., Pasadena, CA 91125, USA}
\begin{document}

\date{Accepted 2017 June 11. ; Received  2017 June 11; in original form 2016 July 12}

\pagerange{\pageref{617}--\pageref{628}} \pubyear{2017}

\maketitle

\label{firstpage}

\begin{abstract}
We report the results of our  radio,  optical and infra-red  studies  of a peculiar  radio source 4C~35.06,
an extended  radio-loud AGN   at the center of  galaxy cluster Abell 407 ($z=0.047$).
The central region of this cluster hosts   a remarkably tight   ensemble of  nine
galaxies, the spectra of which resemble those of passive red ellipticals,  embedded within  a  
diffuse  stellar halo of  $\sim$1~arcmin size.  
This   system  (named  the `Zwicky's Nonet') provides  unique and compelling evidence for 
a multiple-nucleus   cD  galaxy precursor. 
Multifrequency radio observations of 4C~35.06 with  the  Giant Meterwave Radio Telescope (GMRT) at 610, 
235 and 150 MHz reveal a system of  400~kpc scale  helically twisted  and  kinked radio jets and 
outer diffuse  lobes. The outer extremities of  jets  contain  extremely 
steep  spectrum (spectral index  -1.7 to -2.5)  relic/fossil 
radio plasma with a spectral age  of a few$\,\times (10^7 - 10^8)$ yr. 
Such  ultra-steep  spectrum  relic  radio  lobes without definitive hot-spots are rare,  
and they provide  an opportunity  to understand the life-cycle of relativistic jets and  
physics of  black hole  mergers in  dense environments. We  interpret  our  observations of this  radio source in the 
context of the growth of its  central black hole,  triggering of its AGN activity  and  jet  precession,   all possibly 
caused by  galaxy mergers  in  this  dense  galactic system. 
A  slow  conical  precession of the  jet axis due to gravitational perturbation  
between interacting  black holes  is  invoked  to  explain  the  unusual jet morphology.

\end{abstract}

\begin{keywords}
Galaxies: clusters: individual (Abell 407) , galaxies: elliptical and lenticular, cD, galaxies: jets-quasars: supermassive black holes- Radio continuum: galaxies.

\end{keywords}

\section{Introduction}
 Galaxy mergers, which take place more frequently  in  the  dense  environments of galaxy clusters play a pivotal 
 role in the  evolution of galaxies in the Universe across cosmic time. The merger dynamics 
 is influenced by various factors such as  size, mass, impact parameter, relative velocity, 
gas content and the relative  inclination of the participating  galaxies. Mergers have  profound  
effects on   the  properties of galaxies  on  various physical scales. On 
galactic scales ($\sim10 -100$ kpc)  mergers  may result in  ram pressure stripping of gas, 
the formation of long  tidal tails  and  enhanced  star formation. On smaller scales ($< 1$ pc),   
the growth of black holes (BHs) and  possible triggering of AGN activity
 with occasional  relativistic jet ejection may occur  in mergers, 
which   transport  matter and energy from the  galaxy  interiors  to the surroundings   
through  AGN  feedback  processes \citep{MN12}. On the  
 smallest scales ($<<$pc), the  final inspiral stage of  merging BHs  results in 
 powerful gravitational wave emission.  However, the  formation and  merger  
rates of galactic BHs   are  still much  uncertain.  Finally,  on  very large scales ($\sim100 -1000$ kpc), the 
 shocks and turbulence  inducted  into the intra-cluster medium (ICM) during mergers may  
 inject large amounts of  nonthermal  energy and subsequent 
 shock heating of the ICM  to  X-ray temperatures, with disruption of 
 cooling cores \citep{kandu2006,b64,spaul_01}.  Therefore cluster centers are fascinating
laboratories  to  study galaxy formation and  evolution.

\begin{center}
\begin{table*}

  \caption{Galaxy properties in central region of Abell 407 cluster.}
  \begin{tabular}{|c|c|c|c|c|c|c|c|c|c|c|}
  \hline
{\hskip -4.50cm}   Galaxy$^a$  & {\hskip -5.0cm}Coordinates (J2000)& Redshift$^b$ & \multicolumn {5}{c|}{SDSS Magnitudes} & \multicolumn{3}{|c}{SDSS Colours}    
   \\\cline{4-8} \cline{9-11}
  & {\hskip -6.0cm}(from SDSS) &  &$ u$ &$g$ & $r$ &$i$  &$z$ & $g-r$ & $r-i$ & $i-z$\\
 \hline \hline
 {\hskip -4.50cm}G1 (B)& {\hskip -5.0cm}03h 01m 51.5s +35d 50m 30s & 0.0483 & 18.23 & 15.84 & 14.72  & 14.23 &13.84 &1.12 &0.49 &0.39\\
{\hskip -4.50cm}G2 (A)& {\hskip -5.0cm}03h 01m 51.2s +35d 50m 22s & 0.0454 & 21.09 &19.04 & 18.46 & 17.95 & 17.54 &0.58 &0.51 &0.41\\
 {\hskip -4.50cm}G3 (D)& {\hskip -5.0cm}03h 01m 51.8s +35d 50m 20s & 0.0471 & 18.11 &16.00 & 15.00 & 14.44 &13.97 &1.00&0.56 &0.47 \\
 {\hskip -4.50cm}G4 (C) & {\hskip -5.0cm}03h 01m 51.7s +35d 50m 12s & 0.0501 & 20.91 &18.81 & 17.76 & 17.37 &16.90 &1.05 &0.39 &0.47 \\
{\hskip -4.50cm}G5 (F) & {\hskip -5.0cm}03h 01m 51.5s +35d 50m 12s & 0.0478 & 20.02 &17.93 & 16.86 & 16.31 &15.85 &1.07 &0.55 &0.46 \\
{\hskip -4.50cm}G6 (E) & {\hskip -5.0cm}03h 01m 52.4s +35d 50m 29s & 0.0444 & 21.33 &19.93 & 19.11 & 18.70 &18.18 &0.82 &0.41 &0.52 \\
 {\hskip -4.50cm}G7 (G) & {\hskip -5.0cm}03h 01m 53.2s +35d 50m 26s & 0.0471 & 18.37 &16.34 & 15.31 & 14.80 &14.39 &1.03 &0.51 &0.41 \\
{\hskip -4.50cm}G8 (H) & {\hskip -5.0cm}03h 01m 53.7s +35d 50m 28s & - & - &- & - & - &-  &-  & -&\\
{\hskip -4.50cm}G9 (I) & {\hskip -5.0cm}03h 01m 54.5s +35d 50m 18s & 0.0451 & 20.27 &18.68 & 17.59 & 17.23 &16.68 &1.09 &0.36 &0.55\\
\hline
$^a$ Zwicky's original nomenclature is shown  within brackets.\\ 
$^b$  From \cite{b26}.
\end{tabular}
\label{tab1}
\end{table*}
\end{center}



Rich galaxy clusters are  usually dominated  by  massive, luminous  central  elliptical galaxies 
known as  the Brightest  Cluster Galaxies (BCG). There are  some interesting examples, where  extremely 
massive ($> 10^{12} M_{\odot}$) and    brighter ellipticals called the cD galaxies form in the densest   
regions near the spatial and kinematical center of  their host clusters \citep{tov_01,b62}.  The  distinguishing 
property of  cD galaxies   is the presence  of a diffuse, faint stellar halo   that  may extend  
up to  100s of kpc,  well into the  intracluster  medium  \citep{b62,tov_01}. These facts seem to  suggest 
that the formation of cD galaxies is unique to the cluster environment and is  linked closely to 
its dynamical history.   It is still far from clear  how  cD  galaxies
originate in  such a    galactic  environment,  because  very  few  clusters  with  cD galaxies in the
process of  formation  have   been  identified.

 \begin{figure}
 \centering
\includegraphics[width=3.5in,height=4in,keepaspectratio]{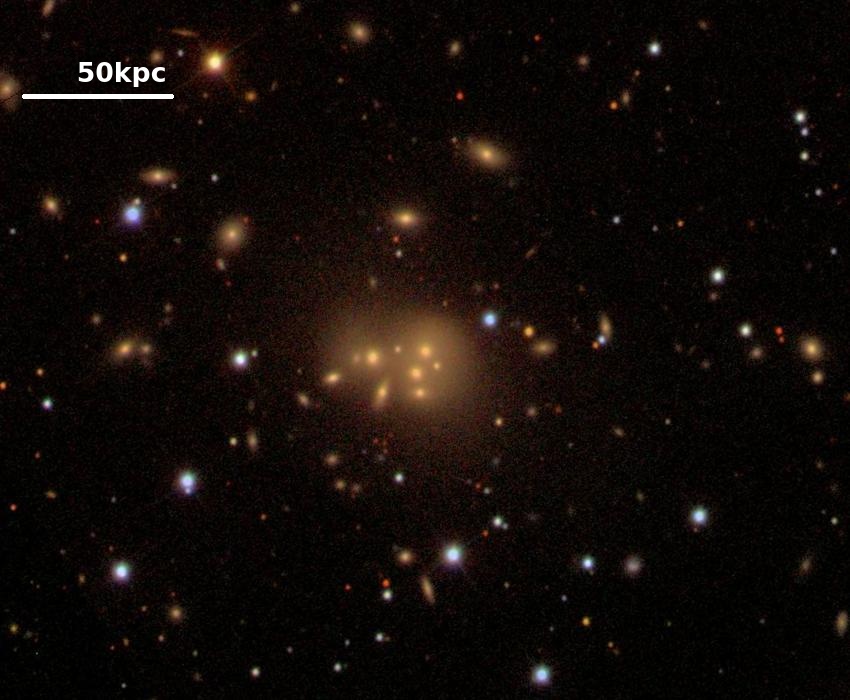}
 \caption{Colour  image ($5.6^{\prime} \times 4.7^{\prime}$) of the  galaxy cluster Abell~407  taken from the Sloan Digital Sky Survey (SDSS). 
  The central region is host to  a   striking   group of  nine  
  close packed  galaxy like condensations,  embedded  within  a diffuse 
 stellar halo of intra-cluster light (the `Zwicky's Nonet').  This  system  possibly represents 
 an exceptionally rare site of  a  multiple-nucleus cD galaxy precursor assembling in a rich galaxy 
 cluster environment. }
\label{fig1}
\end{figure}

\begin{figure*}
 \centering
 \includegraphics[width=3.0in,height=3.5in,keepaspectratio]{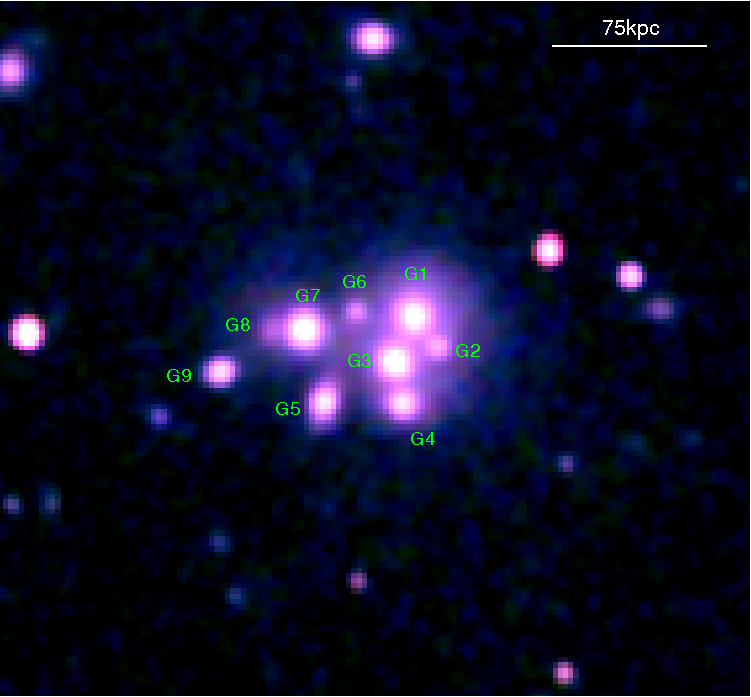}\quad
 \includegraphics[width=2.9in,height=3.5in,keepaspectratio]{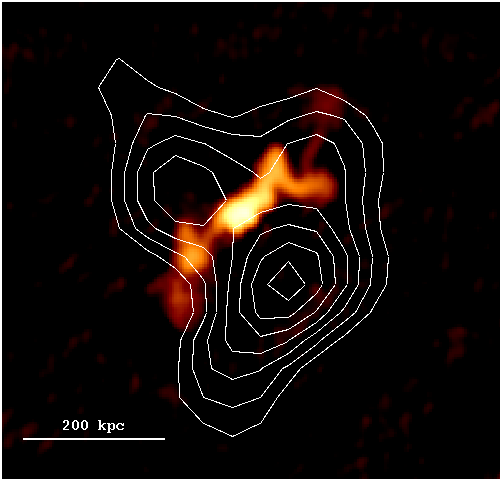}
 
  \caption{Left panel: A pseudo-color near infra-red image ($2.1^{\prime} \times 2.1^{\prime}$) of the compact group `Zwicky's Nonet' 
  obtained  by  combining the J, H and K-band images available in the 2MASS 6X deep survey. 
  The nine central   galaxies  of the  nonet are marked on the image while  Table~\ref{tab1} and  Table~\ref{tabk} 
 show  their  positions, magnitudes, colours  and estimated central 
  black hole masses.  Right Panel: GMRT 150 MHz radio image ($11^{\prime} \times 11^{\prime}$) of 4C 35.06 is shown  
  superposed  on the  ROSAT 0.5 - 2.4 keV band  smoothed X-ray data  shown  with contours.}
  \label{2mass_rass+150}
  \end{figure*}

cD galaxies are almost always  radio loud  and  often 
eject powerful radio jets from accreting supermassive 
black holes \citep{bagchi_94,b64}.  Galaxy interactions and  major mergers 
remove significant amounts of angular momentum 
by gravitational torque  and drive a part of the constituent gas towards  nuclear supermassive
black holes (SMBHs), thereby  triggering the activity of the  central engine  \citep{b72}. 
As a result, cD  galaxies  are   more  likely to  be  radio-loud above 
radio luminosity $\sim 10^{24.5}$ W Hz$^{-1}$  at  1.4 GHz  compared with other  galaxies in a cluster
\citep{bagchi_94}. The multiple  galactic nuclei  observed  in  cDs and BCGs  provide 
 evidence supporting the merger scenario.  In addition,  the SMBHs of 
these galaxies are  believed to  grow  by  multiple  galactic mergers  \citep{volo_2003,kul_2012,b51}. 
Thus,   gravitational perturbations in  the  accretion disc in  the presence of 
closely spaced massive black hole pairs  
may result in the  occurrence of  distorted  radio jets. The  inversion symmetry found in `S' or `Z' 
-shaped radio sources  is ascribed  to the precession of a   spinning black hole \citep{b51,b76} 
and the  associated tilting/perturbation  in the accretion disk \citep{b77,b78}.
The   precessing radio jets  in this situation are likely to trace     
a  characteristic   helical  pattern  on the sky  \citep{b96}. One prominent object showing 
this  rare phenomenon is PKS 2149-158, a
dual  radio-loud elliptical galaxy pair  in the center of cluster Abell 2382,
forming  a  pair of  twisted  jet systems \citep{b1}. The well known `C'-shaped wide-angle tail (WAT) 
source 3C~75 is another striking  example of  twin  AGN  producing  two pairs  of   jets  showing   oscillations
and  interactions \citep{b2}.

In this article, we report radio,  optical and near-IR studies of the extraordinary
radio source 4C 35.06 (B2 0258+35B),  located near the center of the cluster 
 Abell 407, which,  interestingly, also harbors  a remarkably compact  group of  nine
galaxies  embedded  inside  a  diffuse stellar
halo of  faint  intra-cluster light.  We  make  a  detailed study of this remarkable  system,  
which  possibly provides a  unique and compelling  evidence for an ongoing formation  of a giant cD galaxy at the 
cluster center  and   connects  to the  evolution of 
its central black hole  by   mergers.  The radio source  4C 35.06  clearly shows a 
very peculiar  twisted  jet  system  on  100 kpc  scale,  which  we investigate further with  
multi-frequency radio data. 

This paper is organised as follows;  In Section 2 
we discuss the  optical, near-IR, radio  and   X-ray  properties of the system. 
 In Section 3 we discuss the observations and  the data reduction procedure. In Section 4 
 the  main results obtained in the  present study are discussed  in seven subsections. In the 
 last  concluding section, section 5,  we  summarise  our   findings.  
 Throughout this article, a Hubble constant $ H_{0} = 73\,$ $\rm km s^{ -1}\, Mpc^{ -1},$ 
  and cosmological parameters $\Omega_{M} = 0.27$ and $\Omega_{\Lambda} = 0.73$ were used.  
 For  redshift  $z = 0.047$  it  implies the linear scale of  0.885 kpc arcsec$^{ -1}$
  and a luminosity distance of 200 Mpc \citep{b68}. We define the synchrotron emission 
  spectral index  $\alpha$ by  $S(\nu) \propto \nu^{\alpha}$,   where $S(\nu)$  is the  flux density at 
  frequency  $\nu$.

\section[]{`Zwicky's Nonet':  Previous Optical, X-ray and Radio Observations}

Abell~407 is a rich galaxy cluster of Bautz-Morgan class II,  at a redshift of 0.047 \citep{b56}. 
The optical image of the  central region of this cluster  from Sloan Digital Sky Survey (SDSS) shows a complex 
ensemble of at  least  nine galaxy-like  condensations $\sim 1 \arcmin$ across
($\sim50$ kpc),  embedded in  a low surface brightness,  diffuse stellar halo, which is 
reminiscent of a giant  cD galaxy (Figure~\ref{fig1}). Table~\ref{tab1} lists the SDSS optical magnitudes, 
redshifts and colours of these 
nine galaxies.  The  $g-r$ colour index  of $\sim1$ indicates
they are passive,  early-type red  galaxies.  Historically, Fritz Zwicky first noticed  
this extraordinary galactic 
configuration (V Zw 311)  in 1971 \citep{zwicky71}; which was later  
studied in more detail  by \cite{b26} using  the Palomar 200-inch and  60-inch telescopes. 
\cite{b26}  described it as the ``most nightmarish known multiple-nucleus system" and concluded
 that  this  puzzling galactic system possibly represents an  extremely  rare  
and unique snapshot of a giant  cD galaxy caught in its formative stages.  
There were no further  investigations  of  this  highly unusual  galactic  system to understand its nature.  
Here  we propose to name  this extraordinary  galaxy group of nine galaxies as  `Zwicky's Nonet',  honoring 
Fritz Zwicky who first noticed  this  galaxy group.  To our knowledge, this is  the most 
compact and rich system  of multiple galaxies known to date. Some other  famous compact 
groups with multiple members are the `Stephen's Quintet' and 
`Seyfert's Sextet', and the  lesser known  `Zwicky's Triplet' (Arp 103).

\begin{figure*}
 \centering
\includegraphics[width=5.5in,height=5in,keepaspectratio]{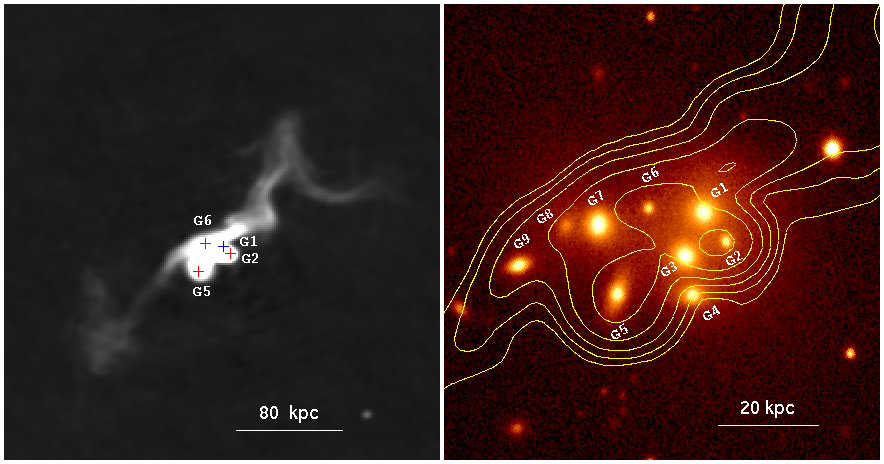}
 \caption{Left panel:  4C~35.06   at  610 MHz   mapped ($5.8^{\prime} \times 5.4^{\prime}$)  with GMRT  at
          $5\arcsec$ resolution,  with the positions of  central  galaxies 
G1 (the brightest member), and G2, G5 and G6 (possible radio sources) marked on it. 
Right panel: the SDSS  zoomed  i-band image ($1.4^{\prime} \times 1.3^{\prime}$) of the  central  region  together  with  the  GMRT  610 MHz  
radio contours plotted on it.  In this figure,  all the nine  galaxies  comprising   `Zwicky's Nonet' 
have  been marked.  North is up, and  east is to the left.}
\label{gmrt610}
\end{figure*}

In the Uppsala General Catalogue,  this  multi-galactic system  is 
 confusingly listed as a  single galaxy UGC~2489  (G1 in our nomenclature) positioned
 at $03^h01^m51.5^s,$ $+35^d50^m30^s$ \citep{b57}. 
 The SDSS optical and  2MASS near-IR images of the central region  of Abell~407 are shown in 
Figs ~\ref{fig1} and \ref{2mass_rass+150} (Left panel), and  a zoomed version is shown in 
Fig. ~\ref{gmrt610} (Right panel), where we have also labeled the  nine galactic  condensations 
by  letters G1 to G9 (see Table~\ref{tab1}). The first column in Table~\ref{tab1} also  gives Zwicky's  
original nomenclature (within brackets) for these nine galaxies.

\subsection{X-ray detection}

 The Abell 407  cluster is   detected in the ROSAT 
 All Sky Survey (RASS), with  estimated X-ray luminosity 
and   gas temperature of  $5 \times10^{44}$ ergs$^{-1}$ (0.1-2.4 keV band) and 2.5 keV 
respectively \citep{ebel_98}. In MCXC (Meta Catalogue of X-ray Galaxy Clusters) it is listed as
MCXC J0301.8+3550  \citep{Piffa2011} with an estimated  cluster mass $M_{500} = 9.16 \times 10^{13}$ M$_{\odot}$, 
where $M_{500}$ is the total mass enclosed inside  a radius $R_{500} = 675.4$ kpc, within which the 
mean over-density of the cluster is 500 times the critical density at the cluster redshift.  
From     archival X-ray  data   in   RASS   hard band  (0.5 -2.4keV)  we  obtained  
surface brightness contours that  are  overlaid on GMRT 150 MHz image (Figure.~\ref{2mass_rass+150}; right panel).  
Here the projected separation of the optical  center  (taken as  G1, the brightest  member)  and the  brighter 
X-ray peak  to the southwest  is $\sim 1.7\arcmin$  or $\sim 90$ kpc. Similar offsets in the 
optical and X-ray emission
peaks have been  observed in  systems  with  ongoing mergers  or in  
dynamically active clusters \citep{ry_08,man_12}.

\begin{figure}
 \centering
 \includegraphics[width=3.5in,height=4in,keepaspectratio]{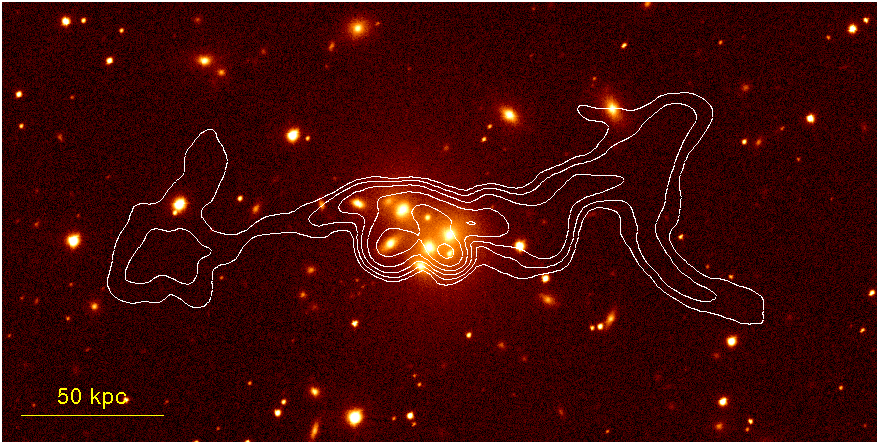}
  \caption{Optical i-band image ($5.7^{\prime} \times 2.9^{\prime}$) of the  galaxy cluster Abell~407  taken from the Sloan Digital Sky Survey (SDSS).
	 The white contours show  the  610 MHz  radio emission morphology of  4C~35.06 as imaged with GMRT at
	 $5\arcsec$ resolution. Both the radio and optical images have been  rotated  for convenience.}
	  \label{610_color}
	  \end{figure}

\subsection{Previous radio observations of 4C~35.06}

  One of the earliest  detections  of this source is at 1.4 GHz frequency with  the 
Cambridge one-mile telescope  with   fairly poor resolution \citep{Riley75}.
Subsequently,  this source  has been studied  using  VLA   at   1.4 GHz and 
5 GHz by \cite{b27}, revealing a bi-lobed structure at $\sim5\arcsec$ resolution. At 1.4 GHz the  
total flux density of  4C~35.06 
is  728 mJy, a core flux density of 10 mJy, an eastern lobe of 305 mJy, 
and a western lobe of 416 mJy. At 5 GHz, the total flux density  was  170 mJy, with core flux of  
4 mJy, an eastern lobe of 55 mJy and a western lobe of 114 mJy. A  5 GHz VLBA  observation at $4.87 \times 2.23$ mas$^{2}$ 
resolution detected a  compact radio core of 2.6 mJy peak flux (3.5 mJy integrated) 
associated with the galaxy G3,  which is  the  second  brightest member of the 
 nonet  \citep{b28}. However, the  VLBA scale is only about 3~pc across, while the large-scale 
radio structure extends  over 200 kpc, leaving a vast gap in our understanding of the  connection 
between the compact AGN   and larger radio morphology.   Moreover,   the previous  
high frequency VLA maps all  miss the large-scale  jet structure and 
 ultra-steep  spectrum   outer regions of this source. Recently, using   high resolution  ($5\arcsec$ FWHM) and 
 high sensitivity (70 $\mu$Jy/beam rms) 
GMRT observations  at 610 MHz frequency,  \cite{b65}  drew  attention to the 
complete  radio  structure  of  an unusual,   helically twisted,   kinked-jet system   
of the radio source 4C~35.06.  \cite{b97}   have  also   studied  this source  
at  a very   low  radio frequency of 62 MHz   with LOFAR  at   angular 
resolution  of  $50\arcsec$ FWHM.  We    discuss  these observations along with  our  new 
GMRT observations in  the  sections  below.

\section[]{Observations and Data Reduction}
\subsection{Multi-wavelength GMRT radio observations}

We  observed 4C~35.06  with  GMRT at  three frequencies; 610, 235 and 150 MHz (Project codes  $21_{-}066$ and   $26_{-}037$ \footnote[1]{https://naps.ncra.tifr.res.in/goa/mt/search/basicSearch}).  Table~\ref{tab2} shows the log  of   radio  observations. For  flux and bandpass calibrations, 3C 48  was  observed at the beginning and end of the  observation  runs  for 10 minutes. 
The 30-minute  scans  on  target source were  alternated  by  5 minute scans  on the phase calibrator. 
 The data were reduced using  NRAO  AIPS software package. AIPS tasks  SETJY and GETJY 
 were used  for  flux density calibrations \citep{b30}. The visibility data were 
flagged for Radio Frequency Interference (RFI) using  AIPS  tasks. The clean and calibrated 
solutions for the flux calibrator were used to calibrate the phase calibrator. The bandpass solutions 
were computed using the  flux  calibrator. These bandpass solutions were  applied to the  data  
and the  frequency channels were averaged to increase the signal to noise ratio. 
The collapsed channel data were recalibrated for phase and amplitude solutions and later applied 
to the target source. AIPS task IMAGR was used for imaging  in 3 dimensions,  
correcting for the   W-term effects  at low frequencies. For imaging, Briggs ROBUST  
weighting parameter  was adjusted to  detect   low surface brightness  
diffuse emission regions better.  Before the final imaging, 
several rounds  of  phase self-calibrations and  one round of amplitude  self-calibration were 
applied  to the data.  At  610 and 235 MHz,  rms  noise levels of  70 $\mu $Jy/beam and 0.90 mJy/beam  
were achieved, respectively. The rms noise measured in  150 MHz image  was  $\sim1$ mJy/beam. 
In  Fig.~\ref{610_color}, the GMRT 610 MHz radio image of 4C~35.06 is 
shown overlaid on the SDSS i-band optical image.

From the VLA archives   we also created  high  frequency radio maps using  data  in  D and C
scaled arrays,  observed  at   6cm (4.8 GHz)  and 20cm  (1.4 GHz) wavelengths, respectively. Standard routines in AIPS were used for calibration and  imaging. For spectral index mapping,  we   imaged both  data  sets with  
identical  $15\arcsec$ (FWHM)  angular resolutions.

\begin{center}
\begin{table*}
 \centering
  \caption{Details of radio observations.}
  \begin{tabular}{@{}llllll@{}}
  \hline\hline
   Telescope&Observed & Band &Obs. time & Beam & Map \\
    &frequency & width & & (arc sec)& rms \\
   
 \hline
  GMRT&610 MHz&32 MHz  & 9hrs & 5.83 $\times $  4.78 &  0.07 mJy/b   \\
  GMRT&235 MHz&6 MHz & 9hrs & 20.86 $\times $16.68 & 0.9 mJy/b    \\
  
  GMRT&150 MHz&16 MHz &  10hrs &19.87 $\times $ 15.77 &  1 mJy/b   \\
  VLA(NVSS)$^a$&1.4 GHz &100 MHz     &survey data  &45 $\times $ 45 & 0.4 mJy/b   \\
  VLA(VLSS)$^b$&74 MHz&1.56 MHz &survey data &80 $\times $ 80  & 100 mJy/b  \\
\hline
$^a$ \citep{b69} & $^b$ \citep{b70} 
\end{tabular}
\label{tab2}
\end{table*}
\end{center}

\begin{figure*}
\includegraphics[width=6.5in,height=6in,keepaspectratio]{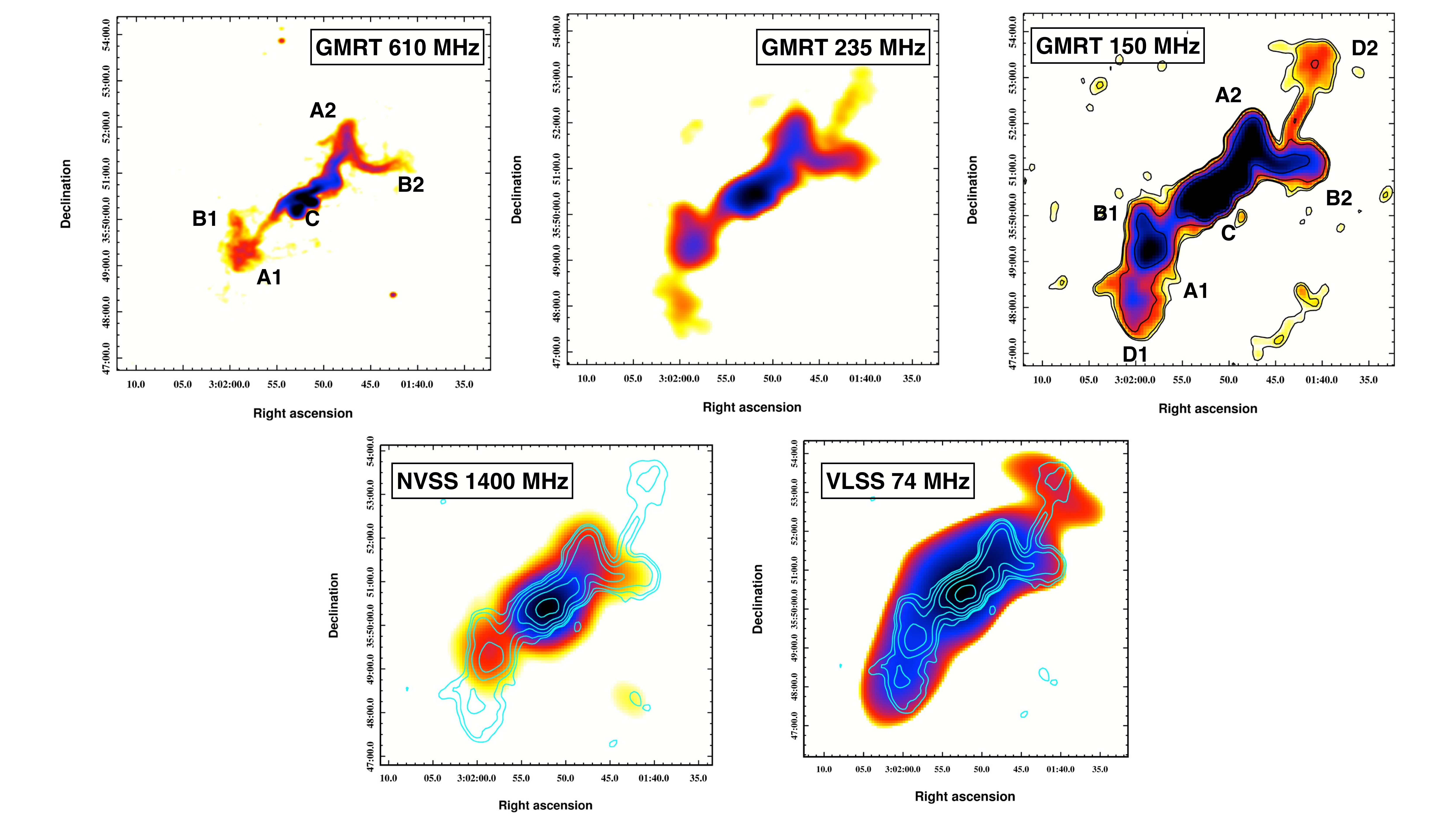}
\caption{Colour scale GMRT images (Top Panel, left to right): at 610 MHz at  resolution $5.83\arcsec 
\times 4.78\arcsec$,  235 MHz at resolution
 $20.86\arcsec \times 16.68\arcsec$, and 150 MHz (contour plot)  with levels
 [1, 2, 4, 8, 16,......]  $ \times 4 \;m$Jy/beam and resolution $23.9\arcsec \times 19.36\arcsec$.
 Bottom Panel: Colour scale images showing 150 MHz GMRT contours plotted over  1400 MHz NVSS
 image (Bottom left ) and  74 MHz  VLSS  image (Bottom right).}
\label{gmrt}
\end{figure*}

\subsection{Spectroscopic  Observations}

We attempted to obtain  good quality spectra for all  nine galaxies comprising  `Zwicky's Nonet'.
The optical spectroscopic observations of the  brighter galaxy members 
G1, G3, G4, G7 and G9 were taken with the IUCAA Girawali Observatory (IGO)  2m telescope 
and fainter  galaxies  G2, G5 and G6  with the
Palomar 200-inch telescope. The aim was to  characterize their AGN and star forming activity, 
and to estimate   the  central velocity dispersions ($\sigma$) and  black hole masses ($M_{BH}$) using the 
well known  $M_{BH} - \sigma$  correlations \citep{b3,b4}.  

Optical long-slit spectroscopic data  were taken   on 20-22 November 2011 on  the IUCAA  2m telescope (IGO).
  The spectra were obtained  using the IUCAA Faint Object Spectrograph and Camera (IFOSC) 
 \footnote[2]{http://igo.iucaa.in}. 
   We used two  IFOSC grisms;  grism no.7 and grism no.8  in combination with a 1.5-arcsec
 slit. These grisms  provide a wavelength coverage of 3800-–6840 \AA\,
 and 5800–-8350 \AA. Standard stars were observed during the same
 nights for flux calibration. Wavelength calibration was done using standard
 Helium-Neon lamp spectra.
Palomar 200-inch observations were carried out on 2014 January 23. The  data were  taken covering 
the  wavelength range 3800 \AA \; to 9500 \AA,  using the double spectrograph (blue and red arms). 
Wavelength calibration was done using standard  Fe-Ar arc lamp  spectra.
 Standard routines in the  Image Reduction  and  Analysis Facility (IRAF)
  were used for data reduction and one-dimensional spectra were extracted  
  using the {\em doslit} task in IRAF. The analysed  data are added as supplementary material. 

\section[]{Results and Discussion}

\subsection{The GMRT images of 4C 35.06:  steep spectrum emission and source  parameters}

Figure~\ref{gmrt} upper row presents the GMRT low frequency radio images at 610, 235 and 150 MHz. 
The regions marked A1, B1 and  A2, B2  represent the features on jets while C 
denotes the core region. The regions D1 and D2 indicate the  outermost diffuse 
structures of the source. The forthcoming sections give a detailed discussion of these regions. 
We  also show  the  radio images from the NRAO VLA Sky Survey (NVSS, 1420 MHz)  and VLA Low-frequency Sky Survey (VLSS, 74 MHz) in the bottom panel of Figure~\ref{gmrt}. 
 The contours of the GMRT  150 MHz image  are  plotted over  the NVSS and VLSS  images for 
size comparison. The highest  resolution (5\arcsec FWHM) and currently the deepest yet GMRT 
image at 610 MHz  shows  a bright,  complex core region and  associated  double 
sided  twisted/helical jet structure,  while the  NVSS 1.4  GHz image  
does  not resolve these structures due to its poor  resolution (45\arcsec FWHM). 
The   VLA  1.4 GHz image  at    $15\arcsec$ resolution    detects  the  extended,    
twisting-turning jet structure  (Fig. ~\ref{gmrt}, bottom left panel). The low  frequency   
235 and 150 MHz  GMRT maps show   extended,  steep spectrum   'relic' plasma  emission features (see Figure~\ref{gmrt})  at  
the extremities  of the helically twisted  jet structure. These features were  also  detected in   
62 MHz LOFAR map [\cite{b97}].  

At 610 MHz the source is found to have a total flux density of $1.7\pm 0.12$ Jy. The flux densities 
along the western and eastern jets are observed to be 580 mJy and 193 mJy respectively. The  angular size 
of the source is $260\arcsec$  or  linear size of $\sim220$ kpc.  The   maximum  extent of the source at 
235 MHz is  $430\arcsec$ or  $\sim380$  kpc. At 150 MHz the source is found to have the largest 
  size of $460\arcsec$  or $\sim400$ kpc linear size. 
This implies that the linear extent of the source   grows larger with the lowering  of  frequency, which is
suggestive of  steep-spectrum  emission regions  present at  the  extremities.
At these frequencies, the western jet is    brighter  than the eastern jet (2.37 Jy and 1.18 Jy at 150 MHz, 
and 1.72 Jy and 800 mJy at 235 MHz respectively). The total  source  flux 
densities at 150 MHz and 235 MHz are  6.0$ \pm$0.18 Jy and 4.7$ \pm$0.13 Jy, respectively. 
 The  excellent quality GMRT images enabled us to  make high resolution ($\sim 25 \arcsec$) spectral 
index maps down to 150 MHz. These maps are  discussed further  in the following sections.

\subsection{Where is the AGN radio core?}

On the  $5\arcsec$ resolution GMRT 610 MHz map  of 4C 35.06, 
  there are  three radio peaks near the center,  of which  two  are  shifted  south  from  
the jet direction (Figure.\ref{gmrt610}). A previous VLBA observation \citep{b28} 
has detected a compact radio  core in the   galaxy G3  on  parsec scale. 
The two  GMRT 610 MHz radio peaks are centered near  optical galaxies  G5 and G2, 
 while the third   radio peak   on the  jet axis is close 
to the faint galaxy G6. Figure~\ref{gmrt610} (left panel) shows the positions of  G2, G5, G6   
and the brightest  galaxy  G1   with `+' signs.  
\cite{b65} suggested that  the  probable  host  AGN    emitting the 
bipolar jet  could be   galaxy   G6 rather than  G3,  which is
clearly  offset  southward from the principal jet direction. The galaxy G6 is 
very faint  both in optical  and  infrared light (Table~\ref{tab1} and~\ref{tabk}).  
From spectroscopy, we   obtained  a  velocity dispersion of 
(143$\pm$40) Km~s$^{-1}$  for G6,  which yielded a  relatively small  black hole mass of 
$(0.52\pm0.65)\times 10^8 M_{\odot}$ (see section 4.7). However, the error margin is  high due to  the 
low SNR of the spectrum. Even though it is  improbable (but not impossible)  for a 
faint galaxy like G6  to  produce such a  large scale radio jet, 
it is   possible  that  this now faint galaxy has been stripped off  the 
majority of its outer halo  stars  in  multiple tidal encounters, 
while  still  retaining a   dense 
stellar core  and  black hole at  the  centre.  Possibly  this is  also 
reflected in  even smaller  black hole
mass,   $(1.5\pm0.07)\times 10^7 M_{\odot}$ derived from its  faint   K-band 
luminosity of $M_K = -21.16$ (Table~\ref{tabk}).  

 In an alternative scenario,  \cite{b97} have  suggested that the rapid  movement of   galaxy G3 
   and its  episodic  AGN radio activity is the reason for the observed peculiar radio  morphology of 
   4C~35.06.  In their interpretation, the  large scale jet morphology  is  due to the 
   earlier phase of activity of  G3, which
  had  switched off  its  radio emission  and  then  restarted while it was  moving to  its  current position,  
 resulting in the offset  inner  double-lobed  morphology   and  steep-spectrum 
 larger jet structure to the  north,  similar to    dying radio
 galaxies \citep{b98}. Thus, at  present, we  are observing  an aged FRI-like large scale structure 
 with an embedded  restarted radio source. 
\cite{b97}  also discuss  that  this AGN core is less likely to be G6 as argued by \cite{b65}, because of the  
 lower mass of this  galaxy and hence the lesser likelihood of it containing an  SMBH.  However, they have 
 not  considered  the   possibility  of stripping of    stars from the 
 outer halo   during  tidal encounters  in  a dense  environment,    
  still retaining the SMBH at its  center. In our opinion, the  canonical   
black hole  mass- IR bulge luminosity  correlation applicable 
for normal ellipticals in relaxed environments may not hold  good in the case of  galaxies   subjected to  violent 
mergers.  We suggest that the observation of the close  coincidence of the optical positions of G2 and G5 
with the radio peaks  and the location of  G6 near the center of symmetry of the large scale jets 
provide ample evidence to argue that the conclusion by \cite{b65}  might be justifiable. 
However, much higher resolution radio  or X-ray  data are still   necessary to  identify the compact  
AGN  core and  the  progenitor  galaxy of the  large scale jet structure firmly.

 \begin{figure}
\centering
\includegraphics[width=3.5in,height=5.0in,keepaspectratio]{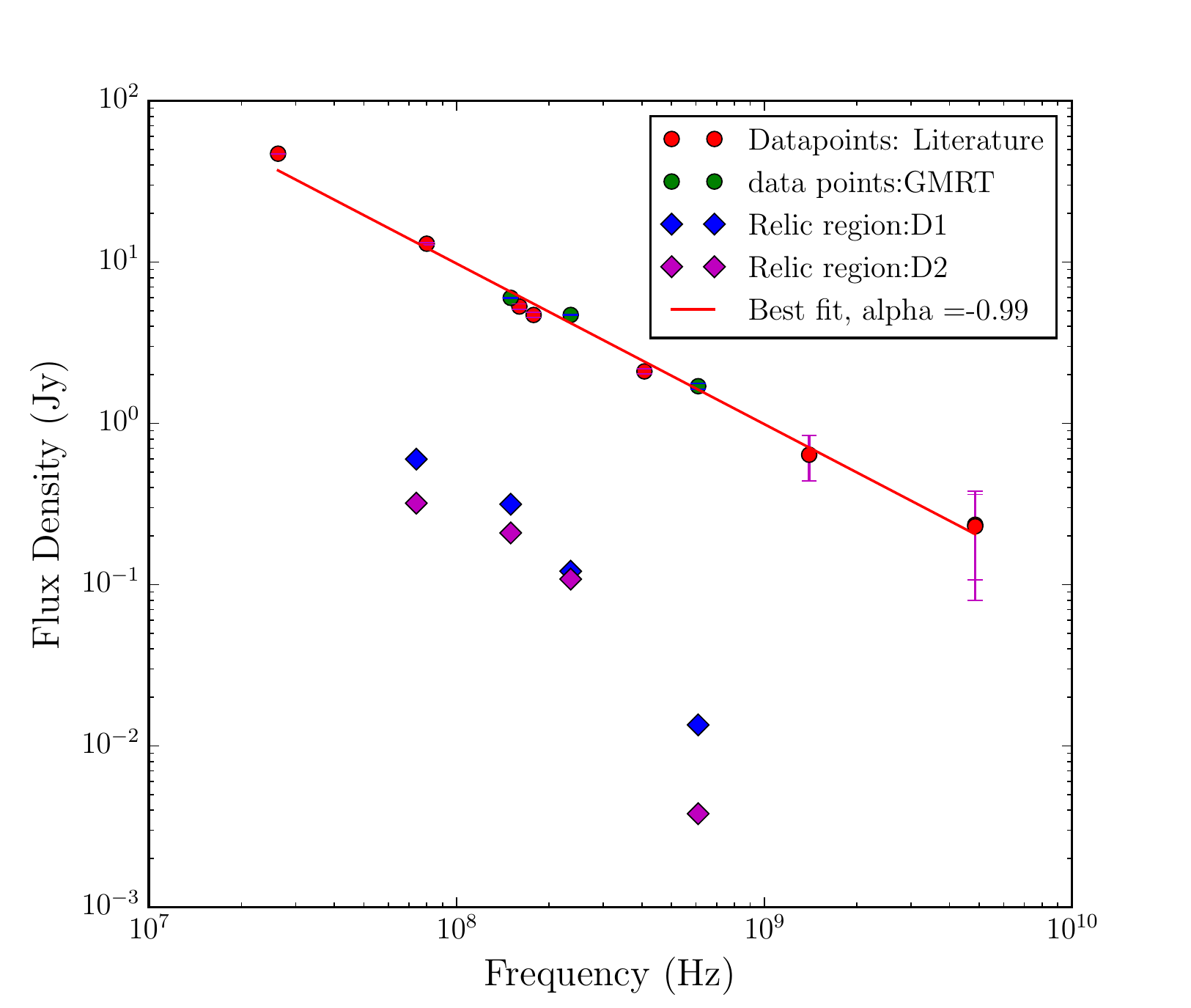}
 \caption{The integrated radio  spectrum of 4C 35.06. A power law fit is shown by a 
 red  line.  Red points are  the data taken from  literature  and green points the GMRT observations. 
 A pair of spectra  shown  with  blue and pink  data points represent the  diffuse,  relic regions D1 and D2 
  detected in GMRT low frequency images (Figure  ~\ref{gmrt}).\vspace{-0.5cm}}
\label{int_spect}
\end{figure}

\subsection{Spectral Index  Maps and  Spectral Ageing Results}


 The  integrated spectrum  and   spectral index maps are  derived from the 
data available in the literature and    our  present  GMRT observations.  
The broad-band   spectrum  shows an  overall  steep  power-law   
from 26 MHz up to 4.9 GHz  (Figure.~\ref{int_spect}). The GMRT  data  points  clearly fit with the 
power-law   spectral index   $\alpha = -0.99$,  thereby  indicating  
 overall  steep spectrum  nature of the source,   as compared to typical radio galaxies with 
jets and  lobes.

\begin{figure*}
 \centering
\includegraphics[width=6in,height=5in,keepaspectratio]{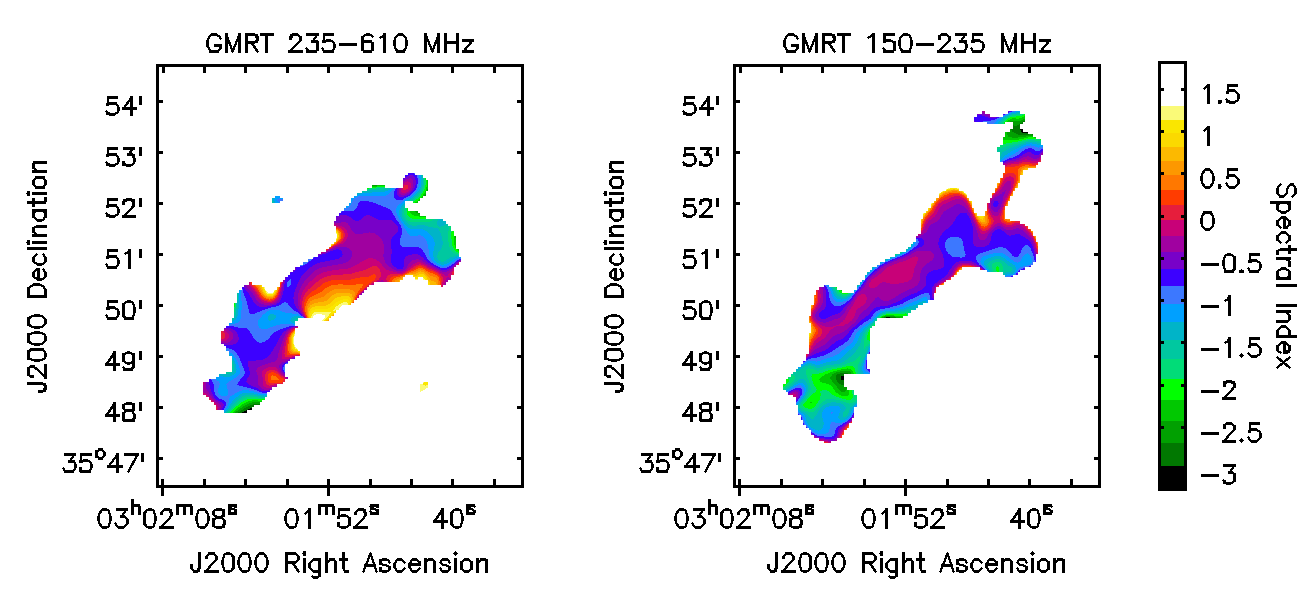}
\caption{The spectral index maps obtained from  235 MHz vs.  610 MHz  (left panel) and 235 MHz vs. 150 
MHz GMRT images (right panel) using matched resolutions. 
The spectral index values are shown 
with a color bar on the right edge. In both  plots, the  spectral index errors  in the 
central region ($\sim 0.02$ and 
$\sim 0.1$ )  are much
lower than that near  the jet extremities  ($\sim 0.2$ and
$\sim 1$ ).
}
\label{spix}
\end{figure*}

To  understand  the  energetics of  this radio source better, we created spectral 
index maps at low and high frequency  ends,  
using radio maps convolved to the   same   resolution. Figure~\ref{spix} shows   dual-frequency 
spectral index maps  of the 235 vs. 610 MHz  and   150 vs. 235 MHz  bands.  The spectral index maps  clearly 
show that the  radio emission in  the central  core  region  has flatter spectra in the range 
  $\alpha = -0.5$  to $ -0.8$, whereas ultra-steep spectrum emission  dominate at  the outer 
extremities,  with   $\alpha \sim  -2$ for 235-610 MHz 
and  150-235 MHz maps. 
The spectral indices for   the  diffuse,  outermost  relic regions marked D1 and D2
are estimated to be   $-1.79$   and  $-2.10$ respectively, indicating  their
ultra steep  spectral  nature (Figure \ref{gmrt} and \ref{int_spect}).
This suggests that the  radio  emission  in region D1 and D2  originates from an
ageing  radio plasma subjected  to heavy energy losses,  possibly  resulting from a
previous  phase  of  energy  injection in the  region. The   LOFAR observations at
62 MHz \citep{b97}  also detect these  ultra-steep spectral regions, but  less  clearly
in comparison with   the    higher sensitivity  GMRT images.
The central region shows a flat spectrum,  possibly due to the superposition  of emission from a few 
 radio-loud AGNs, while the steep spectrum towards the  extremities can be attributed to the lack of 
 fresh injection of accelerated  particles and  radiative  energy losses.

\begin{figure}
 \centering
\includegraphics[width=3.5in,height=5in,keepaspectratio]{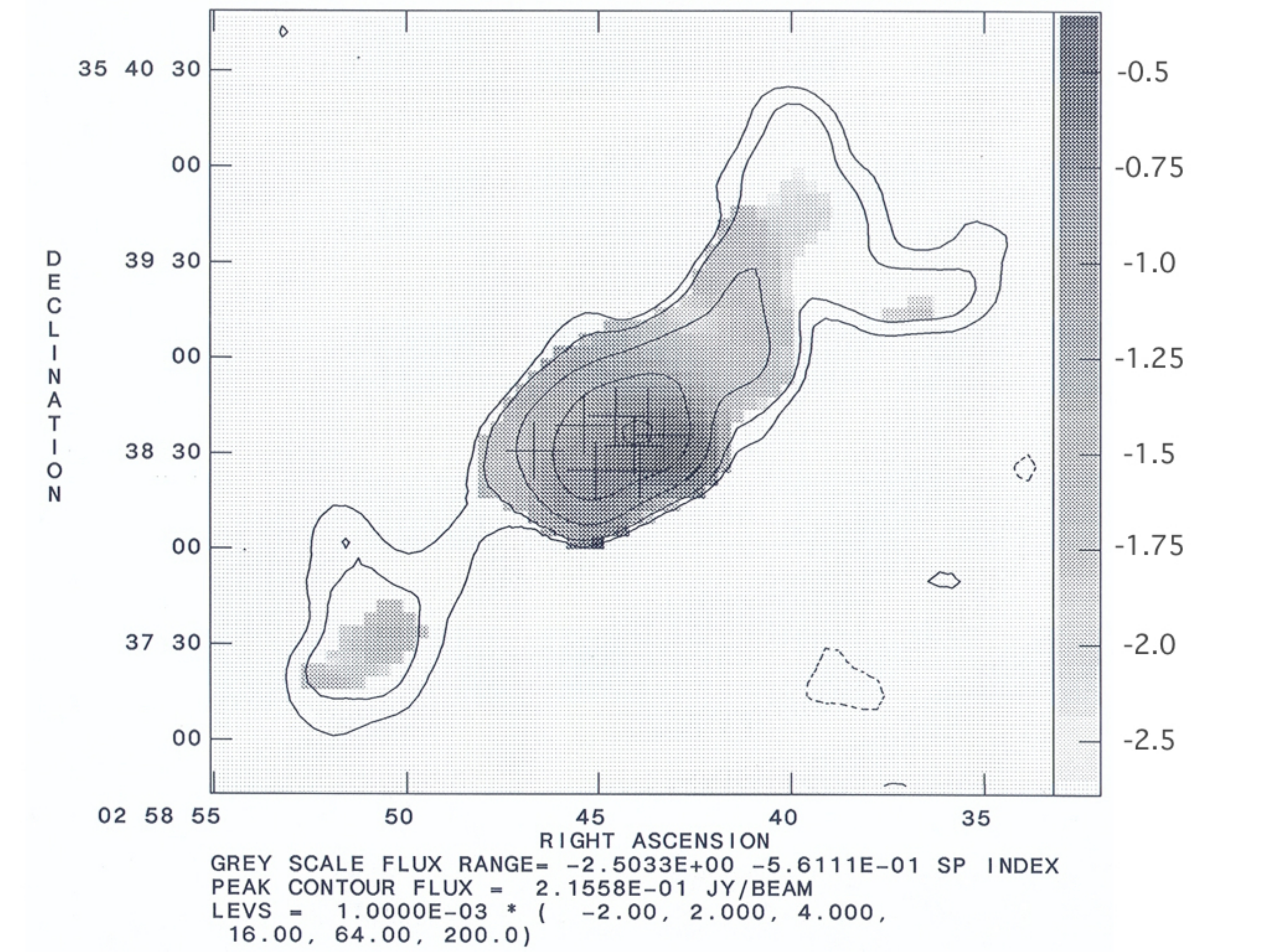}
 \caption{The spectral index map of 4C 35.06  shown in grayscale, obtained by  combining 
 D and C scaled-array   VLA  data at  6cm and 20cm wavelengths  at  $15\arcsec$ resolution.  
 The contours are  the VLA 20cm (1.4 GHz)  radio contours  at the same resolution,  and 
 contour levels  in  mJy/beam  are  shown at the  bottom of the image. 
  The  optical galaxies in the center are denoted with  `+'  symbols.}
\label{vla}
\end{figure}

A high frequency spectral index map (Figure~\ref{vla}) was created by 
combining  VLA  D and C  array  maps   at 6 cm (4.8 GHz) 
and 20 cm  (1.4 GHz)  wavelengths,  both  imaged at the same  $15\arcsec$  angular resolution.  
 This figure   shows  a  flat spectral  index  
central  region  with  $\alpha \approx  -0.5$  (dark shades)  
 which steepens  progressively  away from the center  along the  twisting  radio jets,  
with $\alpha > -2$ (light shades) at the extremities of the jets. Although 
the angular resolution of VLA maps is lower than 
that of GMRT,  the  spectral  index  trend  is  similar to that  with  GMRT maps.

\begin{figure}
 \centering
\includegraphics[width=3in,height=3in,keepaspectratio]{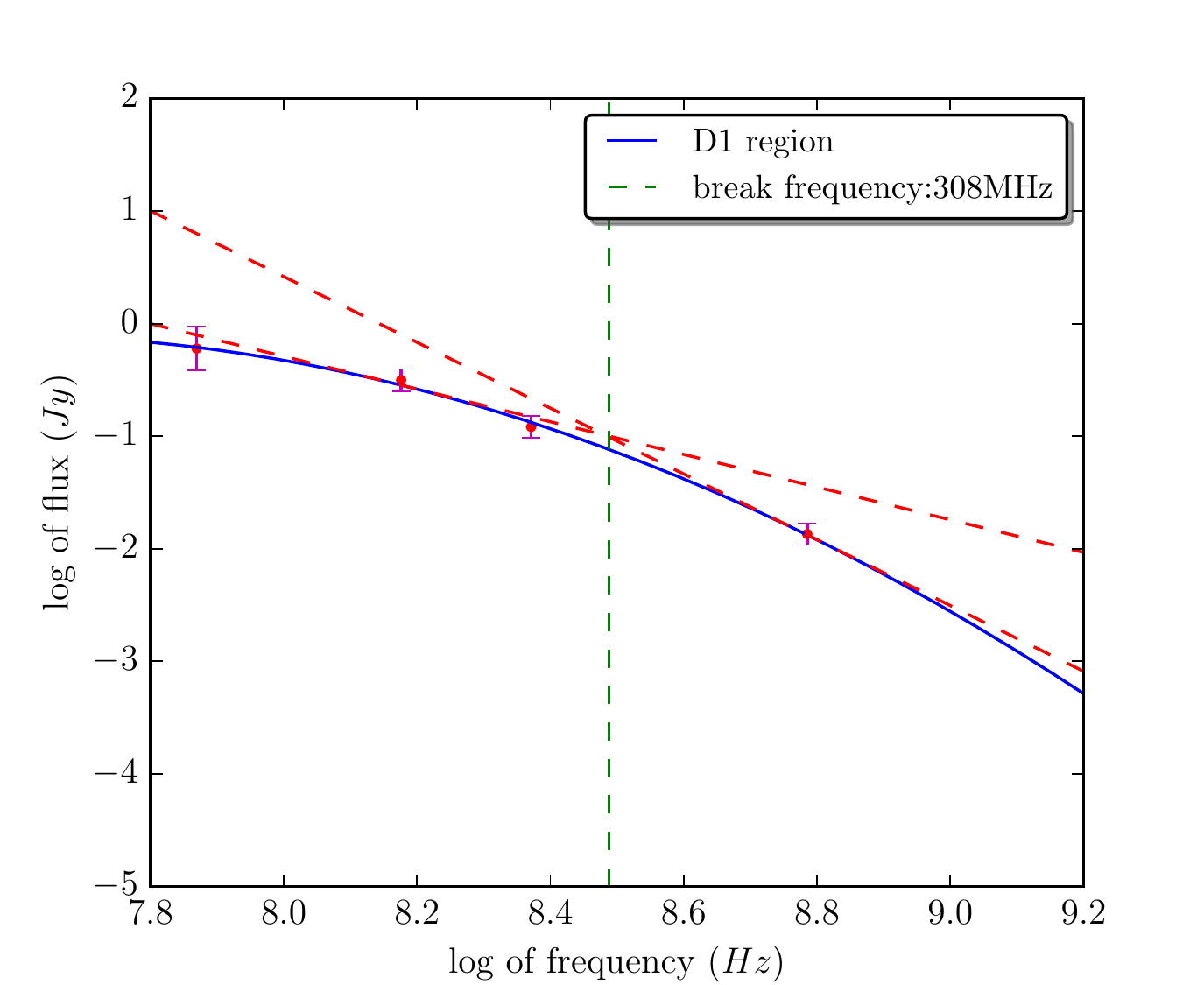}\\
\includegraphics[width=3in,height=3in,keepaspectratio]{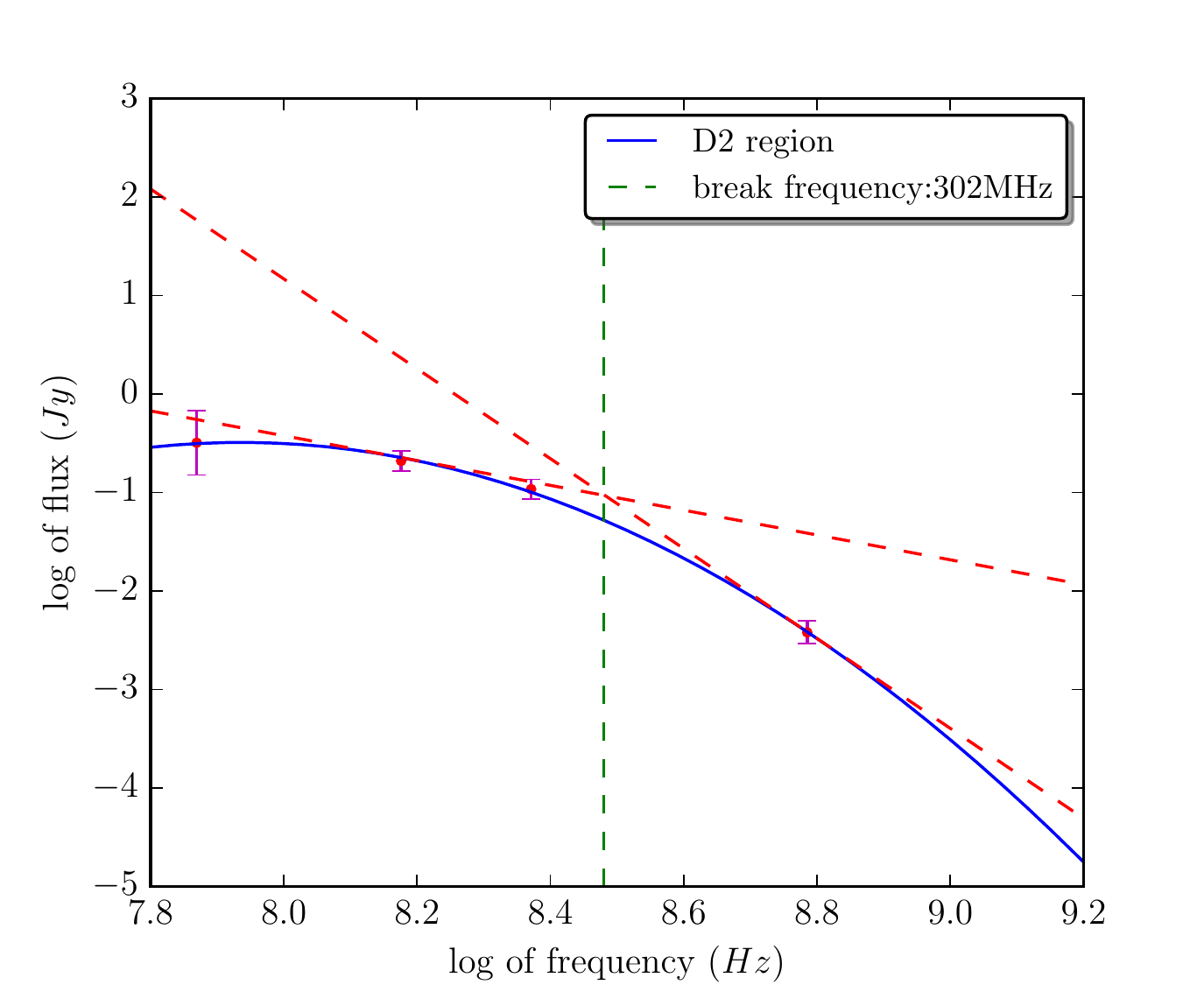}
 \caption{ The second order polynomial  spectral fit (blue line) for the outermost regions D1 (Top panel) and 
D2  (bottom panel).   The  flux values are 
 taken at frequencies 610, 235, 150 (GMRT) and 74 MHz (VLASS). The dashed red lines represent the 
 tangents drawn at higher (610 MHz) and lower frequency (150 MHz) ends of spectra 
 and  the intersection point denotes  the  estimated break frequency (green dotted line).
\vspace{-0.5cm} } 
\label{breaks}
\end{figure}

 The radiative age of  the  non-thermal  plasma in  regions D1 and D2 was  obtained   from  the  spectral 
breaks  in  the   integrated radio spectra  of the regions  shown in  Figure~\ref{breaks}. 
The spectra   are fitted with second order polynomials and  tangents are 
drawn  at   610 MHz  and 150 MHz frequencies. The  intersection point  of these tangents gives
 the break  frequency  $\nu_{b}$,  obtained as 308 MHz and 302 MHz  for  regions
D1 and D2 respectively.  Beyond  the  breaks, a   steepening or a  possible  cut-off  in the  spectra is 
suggested,  consistent with the  scenario  of  radio emission  emanating from
a rapidly cooling  electron population. 

The electron spectral age $t_{sp}$ (or cooling time scale)  is then 
estimated from the synchrotron radio spectrum using the formula  \citep{b98} 

\begin{equation}
t _{sp} = 1.59  \times  10^{9} \left[\dfrac{B^{1/2}}{\left[ B^2+B_{IC}^2\right] \left[\nu _{b}(1+z)\right]^{1/2}}\right] \; yr
\end{equation}

%
This formula is obtained for a uniform magnetic field, neglecting   expansion
losses over the radiative age.  Here  $B$  is the  magnetic field  in $\mu$G, $z$ is the redshift,  
$ B_{IC}= 3.25 (1+z)^2$ $\mu$G  is the inverse Compton equivalent magnetic field and $\nu_{b}$ is the 
cooling break frequency in GHz.
An independent estimate of the magnetic field  
in the relic regions is needed for a robust estimate of  $t_{sp}$.

Assuming minimum energy condition, we calculated the energy density and magnetic field in 
the regions  A1, A2  (the inner loop structures),  and  D1, D2 (the outer relic regions) 
 (see Figure~\ref{gmrt}). The minimum energy density $u_{min}$ is given by:
  \begin{equation}
   u_{min} = \xi(\alpha,\nu_{1},\nu_{2}) (1+k)^{4/7} (\nu_o)^{4\alpha/7} (1+z)^{(12+4\alpha)/7} {(I_o/ d)}^{4/7} 
\end{equation}


where k is the ratio of the energy of the  relativistic protons to that of  electrons, $\alpha$ is the 
spectral index, $\nu_{1}$ and $\nu_{2}$ are lower and
higher limits of frequency, $\nu_o$ is the  frequency at which  surface brightness $I_o$ 
is measured, and function $\xi(\alpha,\nu_{1},\nu_{2})$ is tabulated in  \cite{GF04}. 
Here we  assume the filling factor to be 1,   $\nu_{1}$ = 10 MHz, 
$\nu_{2}$ = 10 GHz, $\nu_o$ = 150 MHz and z = 0.047.

The  equipartition  magnetic field    can be expressed  by:

\begin{equation}
     B_{eq} = (\frac{24\pi}{7} \, u_{min})^{1/2}
\end{equation}

This way the    magnetic fields for the relic regions   are obtained 
as $\approx 5\mu $G and $\approx 16\mu $G 
for  $k =1$ and $k=100$, respectively. The corresponding  elapsed   times
are  $\sim 170$  and $\sim 40$ Myr, respectively. 

We point out that  \cite{b97} calculated the spectral age by estimating the time taken by 
the galaxy  G3 (putative central AGN)  to 
translate from  the  former location (at the  center of the  east-west jet direction), 
by approximating the radial velocity difference between the galaxy and  the stellar envelope  \citep{b26} 
as the velocity of the source in the sky plane. It was assumed that the source was shut off  before 
the translation and restarted once it reached the new position. This approximation enabled 
them to adopt the translation time of G3 to be the shutdown period ($t_{off}$) of the AGN. 
An estimate of spectral age was made  assuming equal time scales   for active ($t_{on}$) and 
quiescent  ($t_{off}$)phases of the AGN and adjudging the sum  (($t_{on}$) + ( $t_{off}$)= 70 Myrs) 
as the age of the  radio plasma. Assuming lowest the Lorentz factor values, they obtained a magnetic field 
of $10 \mu  $G and  a corresponding break frequency of $\sim 380$ MHz, which is  very  close to the  
break frequency ($\sim 300$ MHz)  we   obtained from the  GMRT data points. 
Even though the restarted  AGN activity model and the present model yield almost 
same  values for  break frequency, many assumptions had to be invoked to 
substantiate the former model.\\




\subsection{Twists and  kinks in the jet: Dynamical signatures of  a  perturbed  AGN ?}

\begin{figure}
 \centering
\includegraphics[width=3.5in,keepaspectratio]{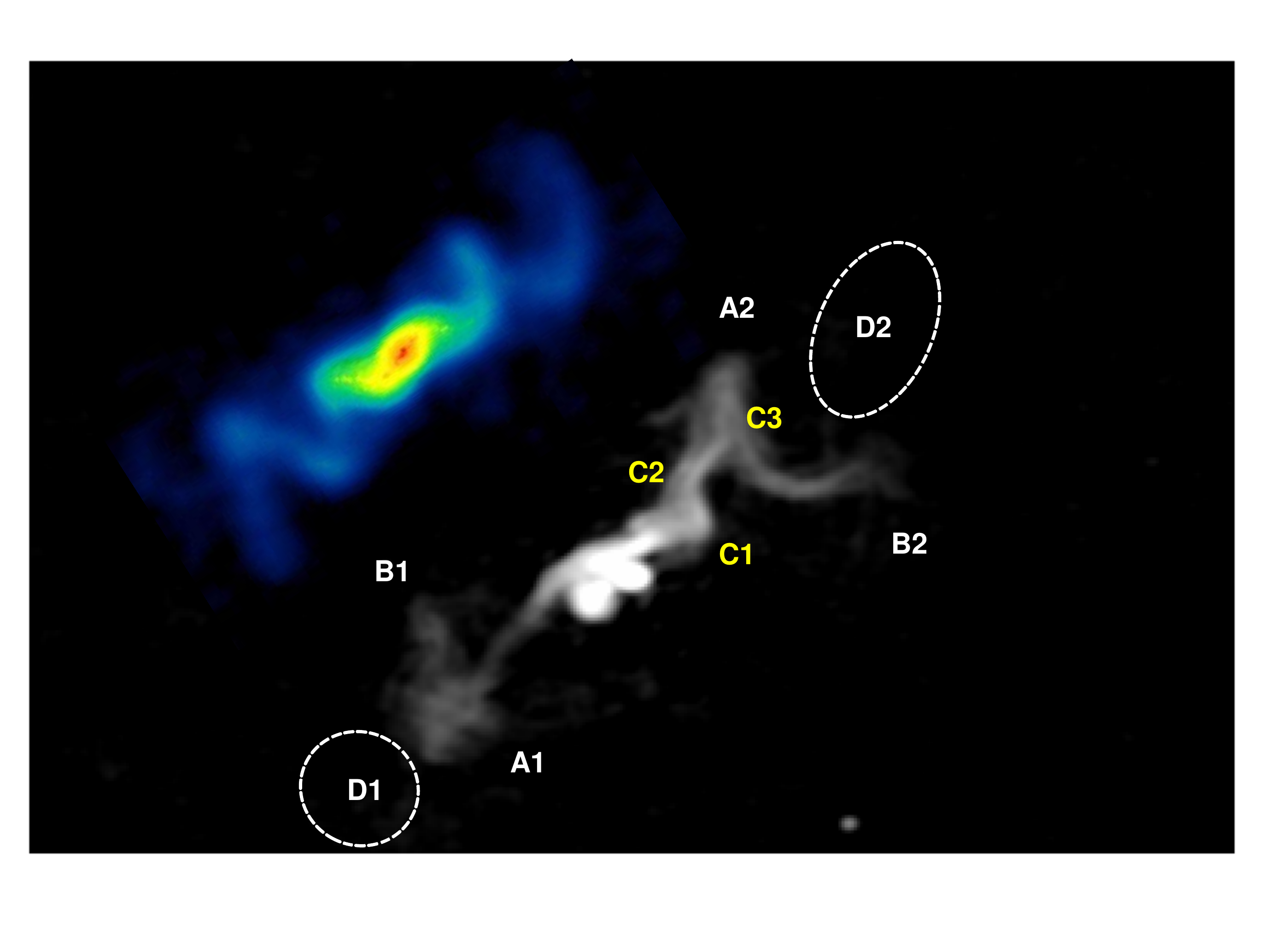}

 \caption{The high resolution (5$^{\prime\prime}$) grayscale image of the source 4C~35.06 at  
 610 MHz (GMRT) showing the twisted, helical jet structure. Different regions of the source are
 marked, and the end extensions D1 and D2  detected   at  235 MHz and 150 MHz,  are indicated 
 by dotted lines.  The colour image shows, for comparison,  the   cork-screw  shaped  precessing  jets observed in  galactic  XRB  `microquasar' SS433, which is the total intensity image at 4.85 GHz 
 \citep{b52}. Note the linear size of SS 433 jet system is 
 only 0.26 pc  while that of 4C 35.06 at  610 MHz  is 230 kpc. }
 \label{ss433}
 \vspace{-0.5cm}
\end{figure}

The  GMRT 610 MHz  maps  (Figures~\ref{gmrt610} and  ~\ref{gmrt}) 
reveal  that the  bipolar  radio  jet  undergoes  helical twists, with  inversion symmetry, 
on either side of   core C  at  the  points  marked   A1 and A2   in Figures.~\ref{gmrt} 
and ~\ref{ss433}.  Moreover, the  north-west arm  of the jet  is observed to be  brighter, 
and bends  to form  a   prominent loop/arc  starting   from    A2  up to   point B2.  A  similarly twisted 
feature  between   points  A1 and B1 is  observed  in  the  south-east arm as well, 
but   fainter and  more  diffuse compared with the north-western  counterpart. 

 The  region D2 in  the western arm  shows  a  peculiar,    upward  bent  jet-like structure in the
150 MHz GMRT map, capped by a    `mushroom' like feature at the top, the nature of which  is not clear at the moment (Figure.~\ref{gmrt} upper panel).  
 The symmetric  counterpart of this   knot/mushroom structure  could be the  downward bent feature  D1  
 in the south-eastern   section of the source. 
 The extremely steep high-frequency radio spectra of the  mushroom like  feature at  D2 (and  also D1),  
 located $\sim$ 200 kpc from the center, indicate  significant   energy  losses,  suggesting  an absence of   freshly 
injected particles. One possibility is that they are  
 buoyant  non-thermal plasma bubbles  rising into the hot   intra-cluster medium, 
  inflated  by  the   radio jet  in the  past, 
 as   observed  around  some   BCG/cD  galaxies   residing in  centers of  clusters or groups   
 \citep{Bagchi09,MN12}. 
  
 In addition to these  interesting  features    there  are  a  few   sharp   kinks  or  steps  in the 
western arm of the jet  at  points marked  C1,  C2 and C3  
(Figure.~\ref{ss433}).   On the eastern side of the jet these  kinks/steps are  not  discernible,  
 possibly due to the projection effects or   their  absence.  
 In our present work, lacking   detailed  modeling,   it is  difficult to  decipher what 
 these  kinks in the jet flow  represent  physically, but they  have been  discussed  
 further  below ( in Section 4.5 $\&$ 4.6).

\subsection{Precessing radio jet structure:  comparison  with galactic  microquasar SS~433}

 The deciphering of the peculiar  radio jet  morphology of   4C 35.06 gains 
added impetus when it is compared with a precessing radio jet structure observed  in the  
galactic `microquasar' SS~433 (see Figure~\ref{ss433}).    
SS~433 is an X-ray binary (XRB) system in the center  of supernova remnant W50,  consisting of a 
 stellar mass black hole or neutron star accreting matter from an A-type supergiant donor 
 star \citep{b31,b32,b52}. 
The most unusual aspect of this object, modelled through radial velocity measurements of 
 `shifting' $H_{\alpha}$ lines and high resolution radio imaging, is that the accretion disk around 
 the compact object  precesses with a regular period of $\sim 164$ days \citep{b31,b32}.
  Consequently, the axis of the jet-ejection nozzle also precesses with the same  period and the
  ejected radio plasma traces out a  dynamically  changing  `corkscrew' pattern
  (see Figure~\ref{ss433}  and  \cite{b52}). 
  
 Even though  a  detailed modelling of the precession geometry of  4C 35.06  is beyond the scope of the 
 present study,  it is  noticeable  that its large-scale jet structure  is  
 analogous to  that of SS 433, if we 
 ignore the  kinks   (C1, C2 and C3) for the time being. The loop portions  
 A2 - B2 and A1 - B1  are suggestive  of 
 the  corkscrew pattern  resulting from   continuous change  of the jet  axis,  possibly 
 due to a precessing  motion.  Further,  flat  spectrum  terminal  hot spots  are absent 
 at  the  jet extremities of 4C  35.06 as well as SS~433,  which is  another indication  of  the  
continuous shifting of the  jet direction.

   Moreover, the  brightness  of the western arm of the 
 jet system  in 4C~35.06 is nearly double that  of the eastern arm.  Depending on the 
 inclination of the jet axis with our line of sight, the  precession cone angle  and the  
 plasma  bulk velocity, the  projected  radio morphology  and  brightness  can appear to be  
 quite different on the approaching and receding sides. This has been  shown  by \cite{b96} in  
 numerical simulation  of  relativistic effects in  precessing jets.
 The differences  in the  observed jet morphology on the two sides can  be attributed as partly due to this.

The precession of  radio jets may be attributable to two mechanisms:  first,   the presence of a 
binary black hole system \citep{b76,b51}, where the torque exerted by the companion black hole 
can precess the   accretion disk of the  first object, leading to  jet precession and secondly,   the Lense-Thirring frame dragging effect  \citep{bp75};   if the angular momentum vector of the accretion disk is  misaligned with 
that of   a  fast spinning Kerr black hole, the black hole  will try to frame drag the inner accretion  
disk so as to align  it  with its  spin vector.  This will lead to the  precession of the accretion 
disk and of  radio jets orthogonal to it. If this  is the reason for the 
helically twisted jets in 4C~35.06, an interesting corollary is that
 the  mass accreting  supermassive black hole  must be spinning. Moreover, the observed  resemblance of
 the morphology of 4C 35.06 to the precessing relativistic jet system of  X-ray binary  SS~433  
 supports the fundamental paradigm that, in spite
 of a vast difference in involved black hole masses, length and time-scales, almost all
 relativistic disk-jet coupled phenomena happen in a scale-invariant 
manner in radio-loud AGNs and  the galactic microquasars. 

\subsection{Jet energetics and  interaction of jets  with the ambient intracluster medium}

In  Figure~\ref{2mass_rass+150} (right panel),   the  GMRT  150 MHz radio image of 4C 35.06 
  is shown  superposed on the   ROSAT  X-ray map in the
  0.5 - 2.4  keV band. Here we  observe that the energetic jet system  of 4C 35.06 
  has  expanded  preferentially in the direction of lower gas pressure in the 
  dense intra-cluster medium.  This might  affect   the ambient  X-ray  medium  by 
  uplifting   the  gas   along the   mean  jet flow  direction.  
  This radio  jet  feedback effect could  be  analogous to  
  the distorted,  extended  lobes  of  supernova  remnant W50  (the `Manatee' nebula), which are  
    shaped by the  interaction of  powerful jets  in  microquasar 
  SS~433 with   the  ambient  ISM   \citep{Dubner98}. 
  We need much deeper X-ray observations of the A407 cluster to   investigate   the AGN feedback 
signatures better.

 Observationally,  the kinetic power of  a  jet ($\bar{Q}$) is a key descriptor of
 the state of  an accreting SMBH system: its mass, spin and the magnetic field of the accretion disk. 
Correlation of low-frequency ( $\nu \sim150$~MHz)  radiative  power   
of radio sources  against their jet power shows  that the radio luminosity of the jet 
constitutes only a small fraction ($<1\%$)
of the total  kinetic power \citep{b66,Daly2012}.
Using the 150 MHz radio flux density  of $6.0 \pm$0.18 Jy from the GMRT map,
the  time averaged kinetic power of jets in 4C 35.06 is computed as  $\bar{Q} \approx 3 \times 10^{43}$erg~s$^{-1}$ \citep{b66}. We have not corrected for the (unknown) loss of energy in the outer 
radio lobes and
thus $\bar{Q}$ is likely to be a lower limit. This power  is below  the  
transition value $5.0 \times 10^{43}$ erg s$^{-1}$  between FR I and II classes. However,
  if the jet continues to operate between $10^{7} - 10^{8}$ yr, the injected  mechanical energy  is 
  $\sim10^{58} - 10^{59}$ erg, which is large enough to  affect or even quench any cooling flow strongly
  and to drive large-scale outflows that redistribute and heat the gas on cluster-wide scales. 
  If we further assume that this  jet power is derived from accretion flow
  onto a  black hole at the rate  $\dot M$ and $\bar{Q} \sim 0.1 {\dot M} c^{2}$, we obtain 
  ${\dot M} \sim 5.3 \times 10^{-3}$ M$_{\odot}$ yr$^{-1}$. This number is only representative, but 
  it suggests  accretion at  sub-Eddington rate $\lambda = \bar{Q}/L_{edd} = 0.0024 \times (10^{8}/M_{BH})$,
  where $L_{edd}$ is eddington luminosity and $M_{BH}$ is the mass of the black hole. This low accretion rate  
signifies  a  radiatively inefficient accretion  flow  (RIAF) in  a  low-luminosity active 
 galactic nuclei (so called LLAGN or LINER). The optical  spectra of  the  galaxies  in Zwicky's 
Nonet (refer Section. 4.7) also confirms this nature.   However    a  
  complete  picture of the launch  of  radio jet in 4C~35.06 and its  energetic feedback 
  effects  requires  much deeper  and higher resolution X-ray  
  and radio data. 

Systems hosting helically modulated 
symmetric jets are the most promising  sites for  finding  close black hole systems 
(triple or binary)\citep{b51,b76}. In Zwicky's Nonet, it is observed that seven galaxy 
pairs  are separated by distances  $\; \sim$ 10 kpc in projection.  Their redshift values are also 
close, with a  mean  $\bar z = 0.0469$ and standard   deviation $\sigma_{z} = 0.00176$, 
or $v \sim 520$ km s$^{-1}$  (refer Table~\ref{tab1}).  The number of binary or triple 
black hole systems discovered with projected separations less than a few kpc are very low \citep{b51}. In the 
present dense system  of nine galaxies,  all packed within a radius of only 25 kpc, 
the smallest projected separation between galaxy pair combinations is about 5 kpc (between G7 and G8).  
So the extreme closeness of the  galactic members coupled with the 
 helically twisted,   large scale jet structure  highlights   the  prime importance of this  
galaxy group  in the  search of  multiple supermassive black hole  systems and  
their  gravitational and electromagnetic  merger  signatures. 
Moreover,  merging SMBHs  would also lead to an enhanced 
rate of tidal disruption of stars and possible  gravitational wave recoil (slingshot) 
ejection of black holes  from galaxies at speeds in excess of  $1000$ km~s$^{-1}$.  
%
 Even though the symmetric helical pattern observed in the jet structure  might be  explained 
  with precession model, there are a few  anomalies like the  presence of  
  distinct kinks or steps  denoted by C1,  C2  and C3  in  the  north-western arm that pose a challenge.
  The precession model alone may be insufficient to explain these features.
  We  note that  C-shaped  twisted paired jet system in  radio source 
  3C~75  is  associated with a  binary black hole pair  
  separated by only 7~kpc \citep{b2}. Both the jets in 3C~75 
  show prominent wiggled  and kinked structures, which have been modelled as  due to the  
  combined linear and   orbital motion of the bound binary black hole  pair \citep{Yokosawa85}.  
 In 4C~35.05  observed  kinks  could    arise from a similar system,  where a black hole  with  
radio jets  is   orbiting  another one at  high speed and  with  large orbital  eccentricity \citep{Yokosawa85}. 
  However,  in  this model,   we can not easily explain  why these kinks or 
steps are absent in the south-eastern  arm of the jet. 

\subsection{Optical spectroscopic results: AGN signature and black hole mass estimation}
  Previously, \cite{b26}   measured the redshifts  and  
  stellar velocity dispersions ($\sigma$) for a  few  of  galaxies  in Zwicky's nonet covering a wavelength
    range from $3700$  to $5250 \AA$.  
    In our  present  study,  we have obtained  good  S/N spectra 
  over the  wavelength range of $3800 $ to $8500 \AA$ for eight out of  the nine galaxies comprising
   Zwicky's Nonet (The spectra are included as supplementary material). 
    However, attempt to obtain a fair spectrum  of  galaxy  G8 failed, due to it being very faint.  Our main aim  
was to search for the signs of  AGN or star forming activity in the optical 
 spectra of these  galaxies and to estimate their black hole masses from the stellar velocity dispersion.
 The  spectra  of these galaxies resemble those of  passive,  early type red 
 ellipticals, devoid of any major emission lines.  
This is not unusual as  optical emission lines 
 are found to be  absent in many  AGNs   showing  radio emission and  large scale radio jets. 
  It has been  observed  that  many  FRI radio sources in galaxy clusters  are 
 hosted by galaxies showing very  weak or  no optical emission lines \citep{b37,b67}.
\begin{center}
\begin{table}
 \centering
  \caption[caption]{K band absolute  magnitudes and  
   masses of the \\\hspace{\textwidth} 
    SMBHs associated with the nine galaxies of `Zwicky's Nonet'. }
  \begin{tabular}{@{}|c|c|c|@{}}
  \hline
   Source &  K Band absolute &  Mass of the \\
   & magnitude & SMBH. \\
   &    & ($ 10^8 M_{\odot} $) \\
 \hline \hline
 G1 &-23.46$\pm$0.039 &1.13 $\pm$0.30\\
 G2 &-21.99$\pm$0.085 &0.31$\pm$0.12\\
 G3 &-23.49$\pm$0.034 &1.16$\pm$0.30 \\
 G4 &-22.60$\pm$0.028 & 0.53$\pm$0.17\\
 G5 &-22.73$\pm$0.031 &0.60$\pm$0.18 \\
 G6 & -21.16$\pm$0.089 &0.15$\pm$0.07 \\
 G7 &-23.56$\pm$0.031 &1.23$\pm$0.30  \\
 G8 & -20.06$\pm$0.113 &0.06$\pm$0.03 \\
 G9 &--22.50$\pm$0.039 &0.49$\pm$0.16  \\
\hline
\end{tabular}
\label{tabk}
\end{table}
\end{center}
Internal properties of a galaxy, such as mass and accretion rate of a SMBH are better estimated from the nuclear emission lines \citep{b39}.
 The  observed  spectra show  that all the  suspected  radio loud  galaxies (G2, G3, G5 and G6)  
belong to the class of low excitation radio galaxies (LERGs)
\citep{b41,b43}.   LERGs  are mostly  found to be hosted by BCGs  having   
extended cD like light profiles \citep{b67},  similar to what we find in  Zwicky's Nonet. 

The well-known tight correlation  which connects  the  mass of the central black hole $M_{BH}$  
to  the galaxy's  bulge  stellar velocity dispersion $\sigma$  \citep{b3,b4} is given by

\begin{equation}
\log_{10}\left(\dfrac{M_{BH}}{M_{\odot}}\right) = \alpha + \beta \log_{10}\left(\frac{\sigma}{ \text{200 km\,s}^{-1}} \right)
\end{equation}

where $\sigma$ is expressed in km s$^{-1}$.  Here we have used  $\alpha = 8.38$ and 
 $\beta= 4.53$, as derived in \cite{b95}. The  estimated SMBH masses  are tabulated in Table~\ref{tab0}. 
 The black hole  masses  were  also calculated using  the  slightly different  $\alpha $ and $\beta $  values taken 
 from \cite{b5} and \cite{b6}. These masses are  consistent within one sigma limits with the  numbers  given in 
 Table~\ref{tab0}. These  results  show that galaxies G1, G3, G5, G7, and G9  all host supermassive black holes 
 of mass ($M_{BH} \approx few \times 10^{8} \, M_{\odot}$). For the other three  galaxies  G2, G4 and
 G6, the  estimate   of   $M_{BH}$  has  large errors. Interestingly, the most 
 massive black hole of mass $M_{BH} \approx 10^{9} \, M_{\odot}$ resides
 in the galaxy G3, which  showed a radio loud AGN  core in  previous VLBA observations \citep{b28}.

The   galaxy  black hole masses  are  also    calculated from 
 their K-band    magnitudes using   6 times deeper  data   on   cluster A407  available from 2MASS survey. 
 The  equation connecting K-band absolute magnitude ($M_K$),
of bulge  component to the  central  black hole mass given by \cite{graham_2007} is, 
\begin{equation}
\log_{10}\left(\dfrac{M_{BH}}{M_{\odot}}\right) = -0.38(\pm 0.06)\left(M_K+24\right) +8.26(\pm 0.11) 
\end{equation}

Table~\ref{tabk} lists the  black hole masses estimated  with this method  for  the nine   galaxies.
The following  caveats are worth mentioning here: It is unclear whether  the  canonical $M_{BH}$-$\sigma$
relation will suffice  for galaxies in such a hostile environment,  undergoing  violent mergers and   stripping of
stars in multiple tidal encounters. This is  clearly  evidenced by the formation of a  large-scale  
stellar halo of  stripped  matter  in Zwicky's Nonet.
The same concern applies if one were  to obtain $M_{BH}$  from the K-band magnitude of bulge using
 the $M_{BH}$-$M_{K}$ correlation.  Moreover,  effect of the
 gravitational potential of the background  stellar halo  (which is highly dark matter dominated; \citep{b26})  
 and close merging  galaxies on the  bulge stellar velocity  dispersion   of a galaxy  
  are also possible   factors that need to be  accounted for in  black hole mass calculations.  
In this article, we have not attempted to do so. However, for checking  this issue, the last column of 
Table~\ref{tab0}  shows  the ratio of black hole mass,    from the $M_{BH}$-$\sigma$ correlation 
($M_{BH,\sigma}$) to that   obtained from $M_{BH}$-$M_{K}$ method ($M_{BH,K}$).  The  ratio 
$M_{BH,\sigma}/M_{BH,K}$  is  $> 2 $ for  galaxies with well determined black hole masses,  
which  suggests that possibly $M_{BH}$-$M_{K}$ method
gives smaller  black hole masses because of the truncation  of the outer envelope  of galaxies, which 
reduces  their  K-band luminosity. Alternatively, the  black hole masses  from the $M_{BH}$-$\sigma$ relation  are overestimated.  

The  stripped  away matter from the  presently observed nuclei must provide a
large fraction of the total luminosity of  the observed stellar halo. The main parameters of the 
   stellar  halo, which  is  detectable up to  the  r-band surface brightness 
   limit of  $\sim 24$ mag arcsec$^{-2}$  (and possibly beyond), 
quoted  by \cite{b26} are as follows; central mass density $\rho(0) = 0.63\pm 0.25 M_{\odot}$ pc$^{-3}$, 
mass-to-light ratio in r band  $M/L = 90\pm35$, and halo radial velocity dispersion 
$\sigma_{halo} = 610 \pm 200$ km s$^{-1}$.
From this value of   $\sigma_{halo}$  and  taking   halo radius  $r \approx  30\arcsec$ ($\sim 26.5$ kpc), 
we obtain the total dynamical mass of halo  as  $2.2 \times 10^{12}$ M$_{\odot}$, which interestingly   is of the same order as that of a super giant cD galaxy. 
\begin{center}
\begin{table}
 \centering
  \caption[caption]{The redshifts and masses of the SMBHs associated \\\hspace{\textwidth}  
   with the galaxy like condensations in `Zwicky's Nonet'.\\\hspace{\textwidth} 
  The last column shows the ratio of the SMBH black hole masses \\\hspace{\textwidth}
  obtained from stellar velocity dispersions and   K band  magnitudes.} 
  \begin{tabular}{@{}|c|c|c|c|c|@{}}
  \hline
   Galaxy &{\hskip -0.5cm}  Red & Velocity & Mass of the &SMBH\\
   &{\hskip -0.50cm} shift& dispersion & SMBH & mass ratio\\
   & &  $(km s^{-1})$ & ($ 10^8 M_{\odot} $)&$M_{BH,\sigma} /M_{BH,K}$ \\
 \hline \hline
 G1 &{\hskip -0.50cm}0.0473 &222 $\pm$16 &3.88$\pm$ 1.23&2.76\\
 G2 &{\hskip -0.50cm}0.0476 &143 $\pm$27&0.53 $\pm$ 0.44&1.71\\
 G3 &{\hskip -0.50cm}0.0470 &273$\pm$18& $9.83\pm 2.96 $&8.47 \\
 G4 &{\hskip -0.50cm}0.0503 & 135$\pm$33 & 0.40 $\pm$ 0.45&0.76\\
 G5 &{\hskip -0.50cm}0.0476 &230$\pm$8 &4.52$\pm$ 0.74&7.5\\
 G6 &{\hskip -0.50cm}0.0454 &143$\pm$40  &0.52$\pm$ 0.65&3.47\\
 G7 &{\hskip -0.5cm}0.0468 &211$\pm$16  &3.08$\pm$1.00&2.50\\
 G9 &{\hskip -0.50cm}0.0445 &176$\pm$20 &1.34$\pm$0.68 &2.74\\
\hline
\end{tabular}
\label{tab0}
\end{table}
\end{center}

\section{Conclusions}

We have presented  the results of  our  radio, optical and infra-red  observations of  the 
 radio source 4C~35.06,  located   in the central region of the  galaxy cluster  Abell 407. 
 The cluster center hosts  a compact  ensemble of  nine  passive,  red  elliptical  galaxies embedded 
within a  faint,  diffuse stellar halo.  We  proposed to name  this  galactic system `Zwicky's Nonet'.

 GMRT observations at 150, 235 and 610 MHz  clearly reveal the 
 complete  radio structure of 4C~35.06, with a  complex  central core  region  and   
helically twisted  and  kinked bipolar radio jets   extending up to $\sim 400$~kpc. The radio jets    
terminate  into  diffuse, ultra-steep spectrum  `relic/fossil'   plasma  lobes  D1 and  D2. In D2, a 
peculiar, very  steep spectrum  ($\alpha < -2$)  mushroom like feature  is  
discovered from  GMRT  150 MHz map.

In regions D1 and D2 of 4C~35.06, the  average minimum energy magnetic field is 
  $B \sim 5$~$\mu$G for  $k =1$ and $B \sim 16$~$\mu$G for  $k =100$. The  corresponding spectral 
ages  of electrons are  obtained as  $ 170 \times 10^6$ and $ 40 \times 10^6$ yr respectively. The  time 
averaged kinetic power of jets  is estimated to be  $\approx 3 \times 10^{43}$erg~s$^{-1}$, indicating that
 the source is a FR I type radio galaxy.

The unique helical jet  system  and  the  very compact  configuration  of nine galactic nuclei point to  
the  possibility of  precessional  and  orbital motion  of the  AGN.  This also  suggests  possible  
gravitational perturbation effects  of   multiple  black  holes residing  in  the extremely dense 
central region of the  cluster.  In such an environment,  orbital  decay  assisted by dynamical 
friction causes the central binary black holes of galaxies to merge, while gravitational torque in the 
binary phase may cause the accretion  disk of AGN to precess,  resulting in a helical jet pattern.   

The  morphological similarity of this  jet system   with that of the   galactic   microquasar SS 433 
also supports  a  precessional scenario. The absence of  terminal  hot spots and  presence of ultra-steep spectrum  
regions on both  ends  of the jet   strongly   suggest  the  continuous shifting of the  jet  direction,  
further supporting the  precessional  model.

  Our study points to the possibility of the fainter member (G6)  of the Zwicky's Nonet 
 hosting   large-scale radio jets. The faintness of this galaxy is  attributed to the  
stripping  of its major stellar envelope   due to the tidal interactions in galactic mergers, 
 retaining the SMBH at the center. The high ratio ($> 2 $) of black hole masses   
 from stellar velocity dispersion and K-band luminosity, i.e.  $M_{BH,\sigma}/M_{BH,K}$, for  
 galaxies with well determined  black hole masses corroborates the diminution of the stellar envelope.
  The  observation of  a  diffuse  stellar halo  of  stripped matter  in the  system  supports this 
   scenario.

The  optical spectra of eight   galaxies  in  Zwicky's Nonet  fail to  show  any prominent emission lines,
  indicating  a  radiatively  inefficient accretion  flow onto the 
central black holes  at  sub-Eddington rates. No strong star-formation/star-burst 
activity  detected    in any of  the  galaxy  spectra. 

Further high sensitivity and  higher resolution radio  observations  are needed  to provide a 
complete spectral analysis and to obtain the  detailed resolved central morphology   of this complex source. 
A deep  X-ray observation of the  hot  intra-cluster  
gas  around the   cD  galaxy precursor,  and   detection of the  AGNs  and   their
X-ray spectra  would  be  very  beneficial in deciphering the nature of this puzzling  radio galaxy.

\section*{Acknowledgments}
We thank the staff of  GMRT,  IUCAA/IGO  and Palomar Observatory for their help during the observations. GMRT is run by the 
National Centre for Radio Astrophysics of the Tata Institute of Fundamental Research.
JB, JJ and  PD  acknowledge generous support from the Indo-French Center for the Promotion of Advanced Research
(Centre Franco-Indien pour la Promotion de la Recherche Avan\'{c}ee)
under programme no. 5204-2.  KGB and JJ acknowledge IUCAA’s
support under the Visiting Associate program. KGB gratefully acknowledges 
the support received through Faculty Development Programme of the UGC, India. 
 AAM and SGD acknowledge partial support from the NSF grants AST-1413600 and AST-1518308. MBP gratefully acknowledges support by DST INSPIRE Faculty Scheme, New Delhi.
We have used images and results from SDSS and funding for SDSS has been provided by the 
Alfred P. Sloan Foundation, the participating institutions, the National Science Foundation, and the U.S. Department of Energy's Office of Science. This research has made use of the NASA/IPAC Extragalactic Database (NED) which is operated by the Jet Propulsion Laboratory, California Institute of Technology, under contract with the 
NASA. We have used VLA data. The VLA is a facility of the National Radio Astronomy Observatory (NRAO).
\def\aj{AJ}%
\def\actaa{Acta Astron.}%
\def\araa{ARA\&A}%
\def\apj{ApJ}%
\def\apjl{ApJ}%
\def\apjs{ApJS}%
\def\ao{Appl.~Opt.}%
\def\apss{Ap\&SS}%
\def\aap{A\&A}%
\def\aapr{A\&A~Rev.}%
\def\aaps{A\&AS}%
\def\azh{AZh}%
\def\baas{BAAS}%
\def\bac{Bull. astr. Inst. Czechosl.}
\def\caa{Chinese Astron. Astrophys.}%
\def\cjaa{Chinese J. Astron. Astrophys.}%
\def\icarus{Icarus}%
\def\jcap{J. Cosmology Astropart. Phys.}%
\def\jrasc{JRASC}%
\def\mnras{MNRAS}%
\def\memras{MmRAS}%
\def\na{New A}%
\def\nar{New A Rev.}%
\def\pasa{PASA}%
\def\pra{Phys.~Rev.~A}%
\def\prb{Phys.~Rev.~B}%
\def\prc{Phys.~Rev.~C}%
\def\prd{Phys.~Rev.~D}%
\def\pre{Phys.~Rev.~E}%
\def\prl{Phys.~Rev.~Lett.}%
\def\pasp{PASP}%
\def\pasj{PASJ}%
\def\qjras{QJRAS}
\def\rmxaa{Rev. Mexicana Astron. Astrofis.}%
\def\skytel{S\&T}%
\def\solphys{Sol.~Phys.}%
\def\sovast{Soviet~Ast.}%
\def\ssr{Space~Sci.~Rev.}%
\def\zap{ZAp}%
\def\nat{Nature}%
\def\iaucirc{IAU~Circ.}%
\def\aplett{Astrophys.~Lett.}%
\def\apspr{Astrophys.~Space~Phys.~Res.}%
\def\bain{Bull.~Astron.~Inst.~Netherlands}%
\def\fcp{Fund.~Cosmic~Phys.}%
\def\gca{Geochim.~Cosmochim.~Acta}%
\def\grl{Geophys.~Res.~Lett.}%
\def\jcp{J.~Chem.~Phys.}%
\def\jgr{J.~Geophys.~Res.}%
\def\jqsrt{J.~Quant.~Spec.~Radiat.~Transf.}%
\def\memsai{Mem.~Soc.~Astron.~Italiana}%
\def\nphysa{Nucl.~Phys.~A}%
\def\physrep{Phys.~Rep.}%
\def\physscr{Phys.~Scr}%
\def\planss{Planet.~Space~Sci.}%
\def\procspie{Proc.~SPIED}%
\let\astap=\aap
\let\apjlett=\apjl
\let\apjsupp=\apjs
\let\applopt=\ao

\bibliographystyle{mn2e}	
\bibliography{a407_1.bib}		

\section*{Supplementary material}
\begin{figure*}
 \includegraphics[width=4in,height=3in,angle=270,keepaspectratio]{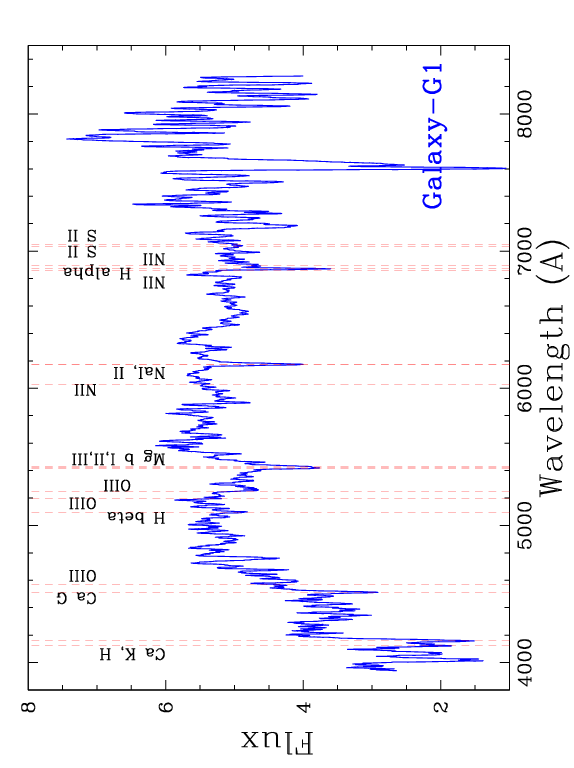} \quad \includegraphics[width=4in,height=3in,angle=270,keepaspectratio]{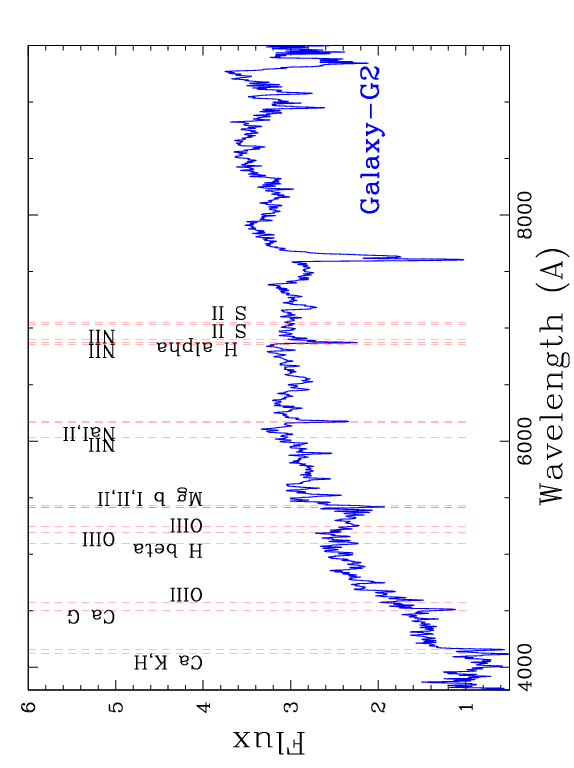} \\
 \includegraphics[width=4in,height=3in,angle=270,keepaspectratio]{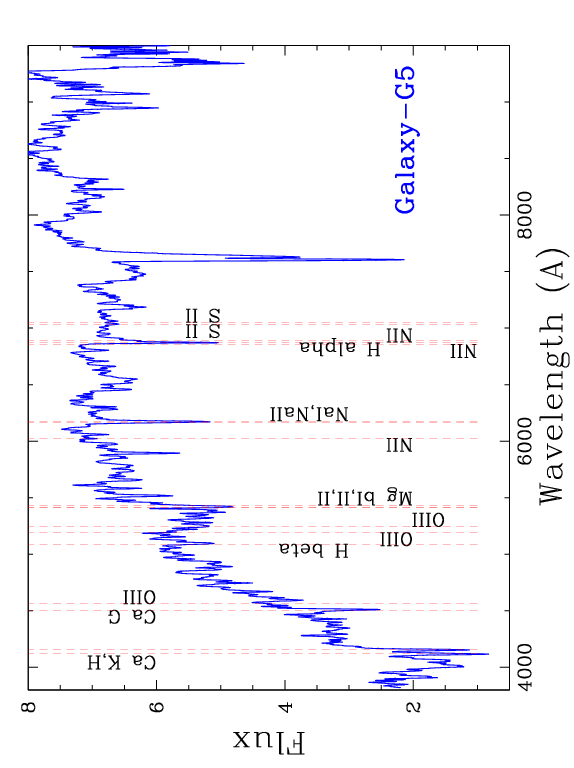} \quad \includegraphics[width=4in,height=3in,angle=270,keepaspectratio]{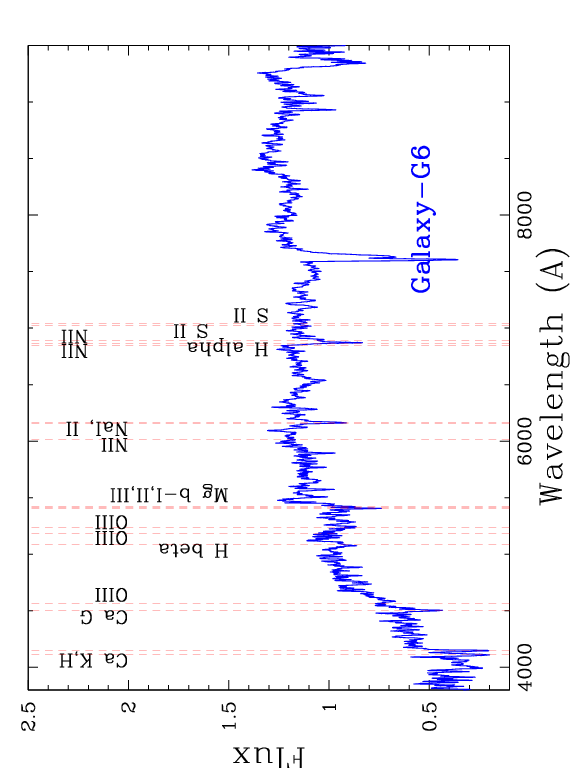}
 
  \caption{The optical spectra of galaxies G1, G2, G5 and G6. The spectra of  members G2, G5 and G6 
  (the galaxies  which are  near the radio intensity peaks of Figure~\ref{gmrt610}) were taken 
  Palomar  200-inch Hale Telescope.  The spectra of the brightest member G1 is taken using the 
  2-meter telescope at IUCAA Girawali Observatory.  Flux scale in   the y-axis is 
  $\times10^{-16}$ ergs cm$^{-2}$ s$^{-1}$ $\AA ^{-1}$.}
\label{fig11}
\end{figure*} 

\begin{figure*}
 \includegraphics[width=4in,height=3in,angle=270,keepaspectratio]{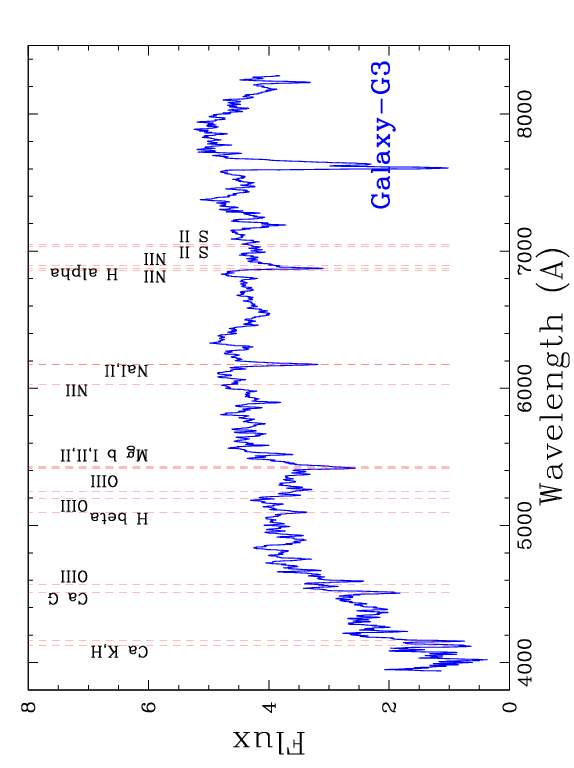} \quad \includegraphics[width=4in,height=3in,angle=270,keepaspectratio]{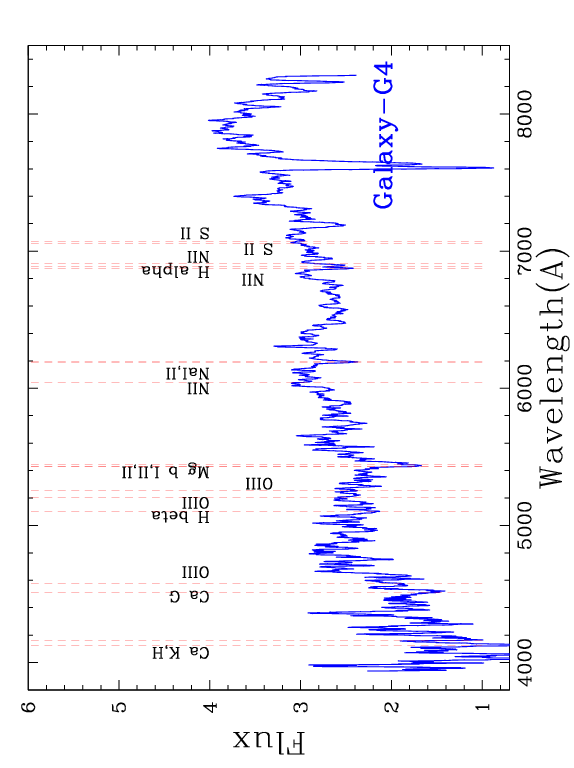} \\
 \includegraphics[width=4in,height=3in,angle=270,keepaspectratio]{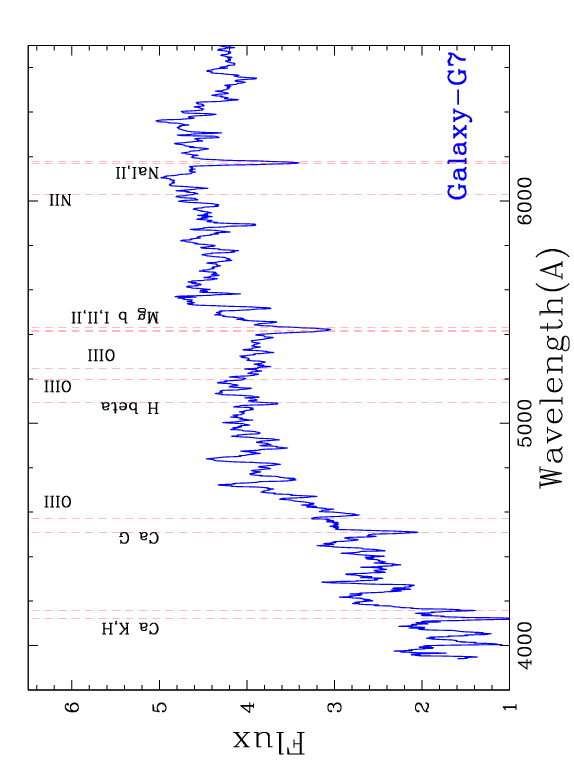} \quad \includegraphics[width=4in,height=3in,angle=270,keepaspectratio]{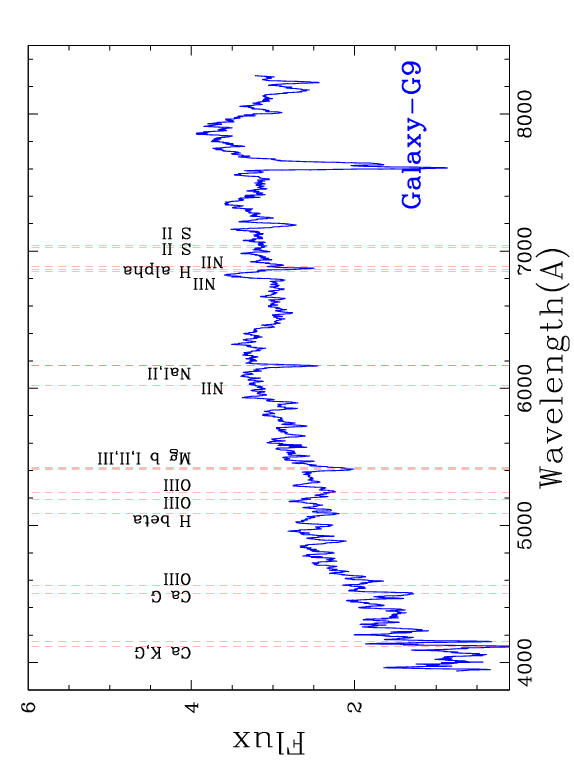}
 
  \caption{ The optical spectra of galaxies G3, G4, G7 and G9 all taken using the 2-meter Telescope at 
  IUCAA Girawali Observatory. Flux scale in  the y-axis is  $\times10^{-16}$ ergs cm$^{-2}$ s$^{-1}$ $\AA ^{-1}$.}
\label{fig12}
\end{figure*}

\label{lastpage}

\end{document}